\documentclass[12pt]{article}
\usepackage{amsmath}
\usepackage{amsfonts}
\usepackage{epsfig}
\renewcommand{\phi}{\varphi}

\newcommand{\beq}{\begin{equation}}
\newcommand{\eeq}{\end{equation}}

\newcommand{\pd}{\partial}

\newcommand{\bphi}{\bar{\phi}}

\newcommand{\tg}{{g}}
\newcommand{\bg}{\bar{g}}
\newcommand{\balpha}{\bar{\alpha}}
\newcommand{\talpha}{\tilde{\alpha}}
\newcommand{\tep}{{\epsilon}}
\newcommand{\ep}{\epsilon}
\newcommand{\lra}{\leftrightarrow}

\title{\bf The reggeon model with the pomeron and odderon:
renormalization group approach}
\author{M.A. Braun$^1$, E.M. Kuzminskii$^{2,1}$, M.I. Vyazovsky$^1$\\
$^1${\small Dept. of High Energy Physics, Saint-Petersburg State University,
St. Petersburg, Russia}\\
$^2${\small Theoretical Physics Division,
Petersburg Nuclear Physics Institute, Gatchina, Russia}}

\begin{document}
\maketitle
{\bf Abstract}

\noindent
The Regge-Gribov model of the pomeron and odderon in the non-trivial
transverse space is studied by the renormalization group technique.
The single loop approximation is adopted.
Five real fixed points are found and the high-energy behaviour of the
propagators is correspondingly calculated. As without odderon, the
asymptotic is modulated by logarithms of energy in certain rational
powers. Movement of coupling constants away from the fixed points
is investigated both analytically (close to the fixed points) and
numerically (far away). In the former case attraction occurs only in
restricted domains of initial coupling constants. More generally in
one third of the cases the coupling constants instead grow large
indicating the breakdown of the single loop approximation.

\section{Introduction}

In  the QCD in the kinematic region
where the energy is much greater than transferred momenta (''the Regge
kinematics'')  strong interactions can be described by the exchange of
pomerons, which can be interpreted  as bound states of pairs of the
so-called reggeized gluons. In the quasi-classical approximation (which
neglects pomeron loops) and in the approximation of a large number of
colors, for the scattering of a  small projectile off a large target
(''dilute-dense scattering'') it leads to the well-known
Balitsky-Kovchegov (BK) equation ~\cite{bal,kov1,kov2} widely used for the description of DIS
and particle-nucleus ($pA$) scattering. The BK equation corresponds
to summing fan diagrams going from the projectile to the nuclear target
with the propagator given by the well-known BFKL equation and the triple
pomeron vertices responsible for the splitting of a pomeron in two ones.
Further generalization to the nucleus-nucleus scattering
in this framework was proposed in ~\cite{braun1a} and studied in ~\cite{braun2a,bond}.
In all cases going beyond the quasi-classical approximation and taking account
of loops presents a hardly surmountable problem, which has been not
solved until now.

In view of this difficulty much attention has been attracted to
 attempts to study the strong interactions outside QCD
within the old reggeon theory \cite{gribov,migdal,migdal1}
introduced by V.N.Gribov and based on the phenomenological local pomeron
and its interaction vertices. The $pA$ interaction in this
framework was considered long ago in \cite{schwimmer} where the sum of all
fan diagrams similar to the BK equation was found.
Unlike the QCD, in the local pomeron model both the pomeron intercept
and coupling constants for the triple pomeron vertex are taken
as phenomenological parameters adjusted to  experimental data.
Needless to say the local pomeron is much poorer in his physical
content as compared to its QCD counterpart, which makes it unfit
to describe processes with hard momentum transfer like DIS.
However, the local pomeron theory is much simpler than the QCD
and admits various methods which make it possible to go beyond
the quasi-classical approximation, the renormalization groups methods in particular.

Still in the realistic two-dimensional transverse world even the local pomeron
model does not allow to find the full quantum-mechanical solution of the
problem with the contribution of pomeron loops fully taken into account.
To overcome this trouble a still simpler model in the zero-dimensional transverse world
(''toy'' model) was considered and studied in some detail
\cite{amati1,amati2,jengo,amati3,rossi,bondarenko,braun3,braun1,braun2}.
Such a model   is essentially equivalent to the standard Quantum
Mechanics and can be studied by its well developed methods.  The important message which
followed from these studies is that 1) the quantum effects, that is the
loops, change cardinally the high-energy behaviour of the amplitudes
and so their neglect is at most a very crude approximation
and 2) passage through the intrcept $\alpha_P(0)=1$ goes
smoothly, without phase transition, to that the theory preserves its physical sense
for the supercritical pomeron with $\alpha_p(0)>1$.

These important findings have been, however, found  wrong in the physical
case of two transverse dimensions. Using the renormalization group technique in
\cite{abarb2} it was concluded that at $\alpha_P(0)=1$ a second order phase transition occurs.
New phases which arise at $\alpha_P(0)>1$ can, however, be hardly considered as physical, since in them
the fundamental symmetry target-projectile is badly violated. So the net result was that the model
cannot accommodate the supercritical pomeron with $\alpha_P(0)>1$ altogether.

It is well-known  that in the QCD, apart from the pomeron with the
positive $C$-parity and signature, a compound state of three reggeized gluons with
the negative $C$-parity and signature, the odderon, appears. Actually it was proposed before the QCD era
on general grounds in ~\cite{nicol}. Since then its  possible experimental
manifestations has been widely discussed
~\cite{odderon1,odderon2,odderon3} with  conclusions containing a large dose of uncertainty up to now, which may be
explained both by the difficulties in the experimental settings and  the elusive properties of the odderon itself.
On the theoretical level in the QCD two species of the odderon
were found, the Bartels-Lipatov-Vacca (BLV) odderon \cite{blv} with the
intercept exactly equal to unity, in which the three reggeized gluons
are pairwise located at the same spatial point, and the more
complicated Janik-Wosiek odderon \cite{jw1,jw2} with all three
reggeized gluons at different points, the intercept somewhat below unity
and so probably subdominant at high energies. It was noted
in \cite{hiim} that the BLV odderon is in a certain sense an imaginary
part of the full $S$-matrix with both $C=\pm 1$ exchanges whose real part
is the pomeron. So the coupling constants for the odderon interactions are
probably the same as for the pomeron interactions.

Following this line of thought in ~\cite{bkv} we introduced the odderon into the
zero-dimensional Regge model to study the influence of the odderon on the properties
of the model. Our numerical results have shown that this influence is minimal.
No phase transition occurs as both intercepts cross unity and the cross-sections
continue to slowly diminish at high energies whether intercepts are smaller or greater than unity.

Of course it is essential to study what occurs in such a model in the two-dimensional
transverse world. Some previous results were obtained in ~\cite{vacca, bartels} within the functional
renormalization group approach,  where two of the five real fixed points were found and
the corresponding structure of the pomeron-odderon interaction was analyzed.

In this note we study the model with the odderon in two transverse dimensions using the renormalization grooup (RG) approach.
We limit ourselves with the lowest non-trivial (single loop) approximation.
We find that presence of the odderon significantly complicates the picture.
Instead of the single attractive fixed point we find many different fixed points which can be complex and are
not fully attractive but remain such
only within a certain reduced domain of initial values. Here we do not discuss all found fixed points but concentrate
on real ones. We first study the behaviour of the propagators in the vicinity of the fixed points
and show that it is qualitatively similar to the case without odderon ~\cite{abarb1} in the sense that at high energies they contain $(\log s)$
in certain rational powers. From the start we put the intercepts for the pomeron and odderon both equal to unity
remembering the situation without odderon.
Of course, one may imagine a possibility that the odderon lifts the singularity at unity intercepts. However, this does not seem
probable to us.
We also study evolution of the coupling constants away from the fixed points. Our results show than roughly in third of the cases the
constants run to infinity indicating that the single loop approximation adopted in our studies is
not sufficient. In the other cases the constants are attracted to three real fixed points.

\section{Model. Renormalization and evolution}

Our model describes two  massless fields $\phi_{1,2}$ for the pomeron $\phi_1$ and odderon $\phi_2$ depending on the rapidity $y$ and acting in the D-dimensional transverse space
with the Lagrangian
\[
{\cal L}=\sum_{i=1}^2\Big(\bphi_{i0}\pd_y\phi_{i0}+\alpha'_{i0}\nabla\bphi_{i0}\nabla\phi_{i0}\Big)\]\beq
+\frac{i}{2}\Big(\lambda_{10}\bphi_{10}(\phi_{10}+\bphi_{10})\phi_{10}+2\lambda_{20}\bphi_{20}\phi_{20}(\bphi_{10}+\phi_{10})
+\lambda_{30}(\bphi_{20}^2\phi_{01}-\phi_{20}^2\bphi_{10})\Big).
\label{eq1}
\eeq
It contains two different slope parameters $\alpha'_{i0}$ for the pomeron and odderon.
Both reggeons are assumed to have zero ''masses'',  $\mu_1=\mu_2=0$, defined as the intercepts minus unity
\[\alpha_i(0)=1+\mu_i,\ \ i=1,2.\]
This choice is related to our desire to apply RG technique. With $\mu<0$ simple perturbation approach is
effective and for $\mu>0$ the theory is badly defined, does not admit direct summation of perturbation series and needs analytic continuation.
As found in ~\cite{abarb2} for the theory without odderon such continuation is prohibited on physical grounds. We postpone investigation of whether
presence of the odderon can improve the situation  for future studies.
The critical dimension of this model is $D=4$ so
the number of dimensions relevant for the application of the
RG technique is $D=4-\ep$ with $\ep\to 0$. Physically of course $D=2$.
This  theory  is invariant under transformation
\beq \phi_1(y,x)\lra \bphi_1(-y.x),\ \ \phi_2(y,x)\lra i\bphi_2(-y,x),
\label{invj4}
\eeq
which reflects the symmetry between the projectile and target.
It has to be supplemented by the external coupling to participants in the form
\beq
{\cal L}_{ext}=i\rho_p(x)\phi(Y/2,x)+i\rho_t(x)\bphi(-Y/2,x)+\rho_p^{(O)}(x)\phi_2(Y/2,x)+i\rho_t^{(O)}(x)\bphi_2(-Y/2,x),
\label{lext}
\eeq
with the amplitude ${\cal A}$  given by
\beq
{\cal A}_{pt}(Y)=-i\Big<{\rm T}\Big\{e^{\int d^2x{\cal L}_{ext}} S_{int}\Big\}\Big> ,
\label{ampli}
\eeq
where $S_{int}$ is the standard $S$ matrix in the interaction representation.
A rather peculiar form for the interaction of the odderon to the participants arises due to specific canonical
transformation of the odderon fields made to simplify its interactions.

We introduce Green functions without external legs
(that is multiplied by the inverse propagator for each leg)
in the energy-momentum representation, which
are  characterized by numbers $m_1,m_2$  and $n_1,n_2$ of reggeons before and
after interaction
\[\Gamma^{n_1,n_2,m_1,m_2}(E, k_i,\alpha_{j0},\lambda_{l0},\Lambda),\ \ j=1,2.\ \ l=1,2,3,\]
where $\Lambda$ is the ultraviolet cutoff.
In the following the superscript $\{n_1,n_2,m_1,m_2\}$ will be suppressed except in special
cases when the concrete numbers $n_i$ and $m_i$ are important.
Our special interest will be in the two inverse propagators
\[\Gamma_1=\Gamma^{1,0,1,0}\ \  {\rm and}\ \ \Gamma_2=\Gamma^{0,1,0,1}.\]
To compare with the model without odderon  ~\cite{abarb2}, if one  introduces $M_i$, $i=1,2$
as the points where the inverse propagators $\Gamma_i(E=M_i,k^2=0)$ vanish, then we require  $M_i=0$.
\beq
\Gamma_i(E,k^2)|_{E=0,k=0}=0.
\label{eq4}
\eeq

Renormalized quantities are introduced in the standard manner:
\[\phi_i=Z_i^{-1/2}\phi_{i0},\ \ i=1,2,\]
\[\lambda_1=W_{1}^{-1}Z_{1}^{3/2}\lambda_{10},\ \ \lambda_{(2,3)}=W_{(2,3)}^{-1}Z_1^{1/2}Z_2\lambda_{(2,3)0}\]
and
\[\alpha'_i=U_{i}^{-1}Z_i\alpha'_{i0},\ \ i=1,2,\]
where we have denoted $W$ the  standard vertex normalization constants
and $U$ new renormalization constants for the slopes.

The generalized vertices transform as
\[\Gamma^{R,n_1,n_2,m_1,m_2}(E_i,k_i,\lambda_i,\alpha'_i, E_N)=Z_1^{\frac{n_1+m_1}{2}}Z_2^{\frac{n_2+m_2}{2}}
\Gamma^{n_1,m_1,n_2,m_2}(E_i,k_i,\lambda_{i0},\alpha'_{i0},\Lambda),\]
where $E_N$ is the renormalization energy point.

Constants $Z$, $U$ and $W$ are determined by the renormalization conditions imposed on renormalizes quantities,
which we borrow from ~\cite{abarb2,abarb1} suitably generalized to include the odderon:

\[\frac{\pd}{\pd E}\Gamma_i^R(E,k^2,\lambda,\alpha', E_N)\Big|_{E=-E_N,k^2=0}=1,\ \ i=1,2,\]\beq
\frac{\pd}{\pd k^2}\Gamma_i^R(E,k^2,\lambda,\alpha', E_N)\Big|_{E=-E_N,k^2=0}=-\alpha'_i,\ \ i=1,2.
\label{renormc}
\eeq
\[\Gamma^{R,1,0,2,0}(E_i,k_i,\lambda_i,\alpha'_j, E_N)\Big|_{E_1=2E_2=2E_3=-E_N, k_i=0}=i\lambda_1 (2\pi)^{-(D+1)/2}.
\ \ i=1,2,3,\ \ j=1,2,\]
\[\Gamma^{R,0,1,1,1}(E_i,k_i,\lambda_i,\alpha'_j, E_N)\Big|_{E_1=2E_2=2E_3=-E_N, k_i=0}=i\lambda_2 (2\pi)^{-(D+1)/2}.
\ \ i=1,2,3,\ \ j=1,2,\]
\[\Gamma^{R,1,0,0,2}(E_i,k_i,\lambda_i,\alpha'_j, E_N)\Big|_{E_1=2E_2=2E_3=-E_N, k_i=0}=i\lambda_3 (2\pi)^{-(D+1)/2}.
\ \ i=1,2,3,\ \ j=1,2.\]
Note that it is necessary to introduce new dimensionless coupling constants $u_0$ and renormalized $u$
\[ u_0\equiv g_{40}=\frac{\alpha'_{20}}{\alpha'_{10}},\ \ u\equiv g_4=\frac{\alpha'_{2}}{\alpha'_1}.\]
The relation between them is determined as
\[ u=u_0\frac{Z_2U_1}{Z_1U_2}\equiv Z_4u_0.\]

With these normalizations the renormalization constants $Z$, $U$ and $W$  depend only on the dimensionless coupling constants
\beq
 g_i=\frac{\lambda_i}{(8\pi\alpha'_1)^{D/4}E_N^{(4-D)/4}},\ \ i=1,2,3\ \ {\rm and}\ \  u\equiv g_4.
 \label{gi}
\eeq

The RG equations  are standardly obtained from the condition that the unrenormalized $\Gamma$ do not depend on $E_N$.
So differentiating $\Gamma^R$ with respect to $E_N$ at $\lambda_{i0}, u_0,\alpha'_{j0}$  fixed we get
\beq
\Big(E_N\frac{\pd}{\pd E_N}+\sum_{i=1}^4\beta_i(g)\frac{\pd}{\pd g_i}+\tau_1(g)\alpha'_1\frac{\pd}{\pd\alpha'_1}
-\sum_{i=1}^2\frac{1}{2}(n_i+m_i)\gamma_i(g)\Big)\Gamma^{R}=0,
\label{eq44}
\eeq
where
\[
\beta_i(g)=E_N\frac{\pd g_i}{\pd E_N},\ \ i=1,...,4,\]
\[\gamma_i(g)=E_N\frac{\pd \ln Z_i}{\pd E_N},\ \ i=1,2,\]
\[\tau_1(g)=E_N\frac{\pd}{\pd E_N}\ln\Big(U_{1}^{-1}Z_1\Big).\]
and the derivatives are taken at $\lambda_{i0}$, $ u_0$  and $\alpha'_{10}$  fixed.

From the dimensional analysis we get
\[
\Big[\Gamma^{R}\Big]=Ek^{D-(n+m)D/2},\ \ n=n_1+n_2,\ \ m=m_1+m_2,\ \ [\alpha'_1]=Ek^{-2}.\]
This allows to write
\beq
\Gamma^{R}(E_i,k_i,g,\alpha'_1,E_N)=E_N\Big(\frac{E_N}{\alpha'_1}\Big)^{(2-n-m)D/4}\Phi
\Big(\frac{E_i}{E_N},\frac{\alpha'_1}{E_N}{\bf k}_i{\bf k}_j, g\Big).
\label{scfun}
\eeq
From this we conclude
\[\Gamma^{R}(\xi E_i,k_i,g,\alpha'_1,E_N)=
\xi\frac{E_N}{\xi}\Big(\frac{E_N/\xi}{\alpha'_1/\xi}\Big)^{(2-n-m)D/4}\Phi
\Big(\frac{E_i}{E_N/\xi},\frac{\alpha'_1/\xi}{E_N/\xi}{\bf k}_i{\bf k}_j, g\Big)
\]\[
=\xi\Gamma^{R}\Big(E_i,k_i,g,\frac{\alpha'_1}{\xi},\frac{E_N}{\xi}\Big).\]

Differentiation  by $\xi$
gives .
\beq
\xi\frac{\pd}{\pd \xi}\Gamma^{R}(\xi E,k^2,g,\alpha'_1,E_N)=
\Big(1-\alpha'_1\frac{\pd}{\pd\alpha'_1}-E_N\frac{\pd}{\pd E_N}\Big)
\Gamma^R(\xi E_i,k_i,g,\alpha'_1,E_N).
\label{eq52}
\eeq

From the RG equation we find
\beq
\Big(\sum_{i=1}^4\beta_i(g)\frac{\pd}{\pd g_i}+\tau_1(g)\alpha'_1\frac{\pd}{\pd\alpha'_1}
-\sum_{i=1}^{2}\frac{1}{2}(n_i+m_i)\gamma_i(g)\Big)\Gamma^{R}=-E_N\frac{\pd}{\pd E_N}\Gamma^{R}.
\label{eq44a}
\eeq
This relation does not change if $E \to\xi E$, so we can
put the left-hand side instead of $-E_N\pd/\pd E_N$ into (\ref{eq52}) and
transferring all terms to the left we find
\beq
\Big\{\xi\frac{\pd}{\pd\xi}-\sum_{i=1}^4\beta_i(g)\frac{\pd}{\pd g_i}+[1-\tau_1(g)]\alpha'_1\frac{\pd}{\pd\alpha'_1}
+\Big[\sum_{i=1}^2\frac{1}{2}(n_i+m_i)\gamma(g)\Big]-1\Big\}\Gamma^R(\xi E_i,k_i,g,\alpha'_1,E_N)=0.
\label{old055}
\eeq
The solution of this equation is standard
\[
\Gamma^R(\xi E_i,{\bf k}_i,g,\alpha',E_N)=\Gamma^R\Big(E_i,{\bf k}_i,
\bg(-t),\balpha'_1(-t),E_N\Big)\]
\beq
\times
\exp\Big\{\int_{-t}^0dt' \Big[1-\frac{1}{2}\sum_{i=1}^2 (n_i+m_i)\gamma_i(\bg(t'))\Big]\Big\},
\label{eq056}
\eeq
where
\beq
\frac{d \bg_i(t)}{dt}=-\beta_i(\bg(t)),\ \ i=1,...,4,
\label{dg}
\eeq
\beq
\frac{d\ln\balpha'_1(t)}{dt}=1-\tau_1(\bg(t)),
\label{dal}
\eeq
with the initial conditions
\beq
\bg(0)=g,\ \ \balpha'_1(0)=\alpha'_1
\label{inic}
\eeq
and
\[t=\ln\xi.\]

\section{Self-masses, anomalous dimensions and $\beta$-functions}
\subsection{Self-masses}

In this study, as mentioned, we restrict ourselves with the lowest order (single loop) approximation.

The unrenormalized inverse propagators have the form
\beq
\Gamma_i(E,k^2)=E-\alpha'_{i0}k^2-\Sigma_i(E,k^2),\ \ i=1,2,
\label{m0}
\eeq
where $\Sigma_i$ are the self-masses. In the lowest approximation they are
graphically shown in Fig.~1.
\begin{figure}
\begin{center}
\epsfig{file=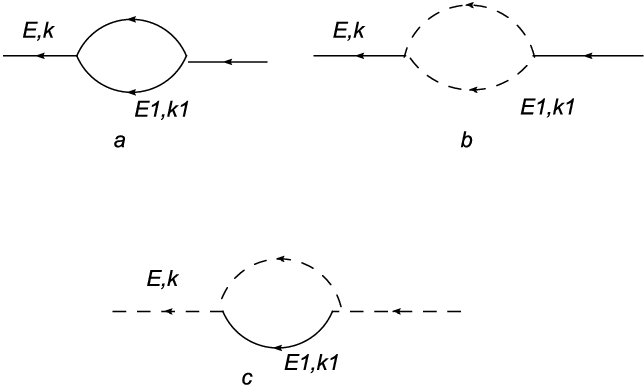, width=14 cm}
\caption{Self masses for $\Gamma_1$ ($a+b$) and $\Gamma_2$ ($c$). Pomerons and odderons are shown by solid and dashed lines, respectively.}
\end{center}
\label{fig1}
\end{figure}
The unrenormalized self- masses are expressed via the unrenormalized parameters $\lambda_{i0}$ and $\alpha'_{i0}$
However, in the lowest order there is no difference between the renormalized and unrenormalized parameters and we use
the former ones.

We start with $\Sigma_1^a$. Explicitly
\[\Sigma_1^a=\frac{1}{2}\lambda_{1}^2\int\frac{dE_1d^Dk_1}{2\pi i(2\pi)^D}\frac{1}{[E_1-\alpha'_{1}k_1^2+i0][E-E_1-\alpha'_{1}(k-k_1)^2+i0]}
\]
\[=-\frac{1}{2}\lambda_{1}^2\int\frac{d^Dk_1}{(2\pi)^D}\frac{1}{E-\alpha_{1}'[k_1^2+(k-k_1)^2]}
=\frac{1}{2}\frac{\lambda_{1}^2}{2\alpha'_{1}}\int\frac{d^Dk_2}{(2\pi)^D}\frac{1}{k_2^2+a^2},\]
where
\[k_2=k_1-\frac{k}{2},\ \ a^2=\frac{1}{4}k^2-\frac{E}{2\alpha'_{1}}.\]
Calculating the integral we find
\[
\Sigma_1^a=\frac{1}{2}\frac{\lambda_{1}^2}{2\alpha'_{1}}\frac{1}{(4\pi)^{D/2}}\Gamma(1-D/2)(a^2)^{D/2-1}
=\frac{1}{2}\frac{\lambda_{1}^2}{(8\pi\alpha'_{1})^{D/2}}\Gamma(1-D/2)\Big(\frac{1}{2}\alpha'_{1}k^2-E\Big)^{D/2-1}.\]
Using  definition (\ref{gi}) of $g_1$
we get finally
\beq
\Sigma_1^a=\frac{1}{2}g_1^2E_N^{2-D/2}\Gamma(1-D/2)\Big(\frac{1}{2}\alpha'_{1}k^2-E\Big)^{D/2-1}.
\label{sig1a}
\eeq
The second part $\Sigma_1^b$ is given by a similar expression with $\lambda_1\to\lambda_3$, $\alpha'_1\to\alpha'_2$ and opposite sign
\beq
\Sigma_1^b=-\frac{1}{2}\frac{g_3^2E_N^{2-D/2}}{u^{D/2}}\Gamma(1-D/2)\Big(\frac{1}{2}\alpha'_2k^2-E\Big)^{D/2-1}.
\label{sig1b}
\eeq
Now the self-mass in $\Gamma_2$ shown in Fig.~1 $c$.
We have
\[\Sigma_2=\lambda_2^2\int\frac{dE_1d^Dk_1}{2\pi i(2\pi)^D}\frac{1}{[E_1-\alpha'_{1}k_1^2+i0][E-E_1-\alpha'_{2}(k-k_1)^2+i0]}
\]\[=-\lambda_2^2\int\frac{d^Dk_1}{(2\pi)^D}\frac{1}{E-\alpha_{1}'k_1^2-\alpha'_2(k-k_1)^2}.\]
In the denominator we find
\[\alpha_1'(1+u)(k_2^2+a^2),\]
where now
\[k_2=k_1-k\frac{u}{1+u},\ \ a^2=\frac{uk^2}{(1+u)^2}-\frac{E}{\alpha'_1(1+u)}.\]
As a result we find
\[\Sigma_2=
\frac{\lambda_2^2}{[4\pi\alpha'_1(1+u)]^{D/2}}\Gamma(1-D/2)\Big(\alpha'_1k^2\frac{u}{1+u}-E\Big)^{D/2-1}.\]
Passing to $g_2$ we have finally
\beq
\Sigma_2=\frac{g_2^2E_N^{2-D/2}}{[(1+u)/2]^{D/2}}\Gamma(1-D/2)\Big(\alpha'_1k^2\frac{u}{1+u}-E\Big)^{D/2-1},
\label{sig2}
\eeq
or in terms of $\alpha'_1$ and $\alpha'_2$
\beq
\Sigma_2=\frac{\lambda_2^2}{[4\pi(\alpha'_1+\alpha'_2)]^{D/2}}
\Gamma(1-D/2)\Big(k^2\frac{\alpha'_1\alpha'_2}{\alpha'_1+\alpha'_2}-E\Big)^{D/2-1}.
\eeq
In the latter form the symmetry between $\alpha'_1$ and $\alpha'_2$ is explicit.

\subsection{Renormalization constants $Z$ and $U$}
To find the anomalous dimensions we have first to find the renormalization constant $Z_1$ and $Z_2$.
They are determined from the normalization conditions (\ref{renormc}).

The renormalized functions $\Gamma_i^R$ are defined as
\beq
\Gamma_i^R=Z_i\Gamma_i=Z_i(E-\alpha'_{i0}k^2-\Sigma_i)=Z_iE-U_{i}\alpha'_{i}k^2-\Sigma_i(E,k^2),
\eeq
where we put $Z_i=1$ in front of $\Sigma_i$ having in mind the lowest nontrivial order.
We can rewrite it as
\[\Gamma_i^R=E-\alpha'_i k^2+\Big((Z_i-1)E-(U_{i}-1)\alpha'_ik^2-\Sigma_i(E,k^2)\Big).\]
The quantity in the bracket is the renormalized mass (with the opposite sign)

\beq
\Sigma_i^R=\Sigma_i-(Z_i-1)E+(U_{i}-1)\alpha'_i k^2,
\label{sigr}
\eeq
 so that
\beq
\Gamma_i^R=E-\alpha'_ik^2-\Sigma^R_i(E,k^2).\label{gamr}\eeq

The renormalization conditions (\ref{renormc}) tell us
\[\frac{\pd}{\pd E}\Sigma_i^R(E,k^2)_{E=-E_N, k=0}=0,\]
which determines
\[Z_i-1=\frac{\pd}{\pd E}\Sigma_i(E,k^2)_{E=-E_N, k=0}.\]

Likewise
\[\frac{\pd}{\pd k^2}\Sigma^R_i(E,k^2)_{E=-E_N. k=0}=0,\]
which gives
\[(U_{i}-1)\alpha'_i=-\frac{\pd}{\pd k^2}\Sigma_i(E,k^2)_{E=-E_N. k=0}.\]

Note that self-mass $\Sigma_1$ is the sum of two contributions from the
pomeron and odderon intermediate states: $\Sigma_1=\Sigma_1^a+\Sigma_1^b$.
So the renormalization constants and anomalous dimensions will be the sum
of contributions $a$ and $b$ from $\Sigma_1^a$ and $\Sigma_1^b$, respectively.

From our formulas for the self-masses we have
\[\frac{\pd}{\pd E}\Sigma_1^a=
-\frac{1}{2} g_1^2 E_N^{2-D/2}\Gamma(1-D/2)\Big(\frac{1}{2}\alpha'_{1}k^2-E\Big)^{D/2-2}
(D/2-1)\]\[=\frac{1}{2}g_1^2E_N^{2-D/2}\Gamma(2-D/2)\Big(\frac{1}{2}\alpha'_{1}k^2-E\Big)^{D/2-2}.\]
At $E=-E_N$ and $k=0$
\[Z_1^a-1=\frac{\pd}{\pd E}\Sigma_1^a|_{E=-R_N,k=0}
=\frac{1}{2}g_1^2\Gamma(2-D/2).\]
Similarly (with obvious expression $Z_1-1=(Z_1^a-1)+(Z_1^b-1)$),
\[Z_1^b-1=\frac{\pd}{\pd E}\Sigma_1^b|_{E=-R_N,k=0}
=-\frac{1}{2}\frac{g_3^2}{u^{D/2}}\Gamma(2-D/2)\]
and finally
\[Z_2-1=\frac{\pd}{\pd E}\Sigma_2|_{E=-R_N,k=0}
=\frac{g_2^2}{[(1+u)/2]^{D/2}}\Gamma(2-D/2).\]

Passing to $U_i$ we find
\[\frac{\pd}{\pd k^2}\Sigma_1^a=
\frac{1}{2}g_1^2E_N^{2-D/2}\frac{1}{2}\alpha'_1\Gamma(1-D/2)\Big(\frac{1}{2}\alpha'_{1}k^2-E\Big)^{D/2-2}
(D/2-1)\]\[=-\frac{1}{4}\alpha'_1g_1^2E_N^{D/2-2}\Gamma(2-D/2)\Big(\frac{1}{2}\alpha'_{1}k^2-E\Big)^{D/2-2},\]
from which it follows
\beq
U_{1}^a-1=\frac{1}{4}g_1^2\Gamma(2-D/2).\label{ua}\eeq
Likewise
\[\frac{\pd}{\pd k^2}\Sigma_1^b
=-\frac{1}{4}\alpha'_2\frac{g_3^2}{u^{D/2}}\Gamma(2-D/2)\Big(\frac{1}{2}\alpha'_{2}k^2-E\Big)^{D/2-2},\]
so that
\beq
U_{1}^b-1=-\frac{1}{4}\frac{g_3^2u}{u^{D/2}}\Gamma(2-D/2).\label{ub}\eeq

Finally
\[\frac{\pd}{\pd k^2}\Sigma_2=
\alpha'_1\frac{u}{1+u}\frac{g_2^2}{[(1+u)/2]^{D/2}}\Gamma(2-D/2)\Big(\frac{u}{1+u}\alpha'_{1}k^2-E\Big)^{D/2-2}\]
and as a result
\beq
U_{2}-1=\frac{1}{1+u}\,\frac{g_2^2}{[(1+u)/2]^{D/2}}\Gamma(2-D/2).\label{u2}\eeq

Knowing $Z$ and $U$ we can write the renormalized masses at arbitrary $E$ and $k^2$.
We find from (\ref{sigr}) and (\ref{sig1a})
\[\Sigma^{Ra}_1(E,k^2)
=\frac{1}{2}g_1^2E_N^{2-D/2}\Gamma(1-D/2)\Big(\frac{1}{2}\alpha'_{1}k^2-E\Big)^{D/2-1}
-\frac{1}{2}g_1^2\Gamma(2-D/2)\Big(E-\frac{1}{2}\alpha'_1k^2\Big).\]
Presenting
\[\Gamma(1-D/2)=\frac{\Gamma(2-D/2)}{1-D/2}\]
we  rewrite it as
\beq
\Sigma^{Ra}_1(E,k^2)
=\frac{1}{2}g_1^2\Gamma(2-D/2)(\alpha'_{1}k^2/2-E)\Big(\frac{((\alpha'_{1}k^2/2-E)/E_N)^{D/2-2}}
{1-D/2}+1\Big) .
\label{sigra}
\eeq
In the similar way we find substituting $g_1\to g_3$, $\alpha'_1\to u\alpha'_1$ and changing sign
\beq
\Sigma^{Rb}_1(E,k^2)
=-\frac{1}{2}\frac{g_3^2}{u^{D/2}}\Gamma(2-D/2)(u\alpha'_{1}k^2/2-E)\Big(\frac{((u\alpha'_{1}k^2/2-E)/E_N)^{D/2-2}}
{1-D/2}+1\Big)
\label{sigrb}
\eeq
and finally using (\ref{sig2})
\[
\Sigma^R_2(E,k^2)\]\beq
=\frac{g_2^2}{[(1+u)/2]^{D/2}}\Gamma(2-D/2)[u\alpha'_{1}k^2/(1+u)-E]\Big(\frac{[(u\alpha'_1k^2/(1+u)-E)/E_N]^{D/2-2}}
{1-D/2}+1\Big).
\label{sigr2}
\eeq

To find the anomalous dimensions we have to differentiate the renormalization constants over $E_N$.
In the lowest order we have for all renormalized constants
\[\frac{\pd}{\pd E_N}\ln Z=\frac{\pd}{\pd E_N}\ln (1+Z-1)=\frac{\pd}{\pd E_N}(Z-1).\]
All renormalized constants depend on $E_N$
via constants $g_i$, $i=1,2,3$ which in the lowest order are equal to the unrenormalized $g_{i0}$
($g_4=u$ does not depend on $E_N$ in this order).
\[ E_N\frac{\pd}{\pd E_N}g_i^2=E_N\frac{\pd}{\pd E_N}\frac{\lambda_i^2}{(8\pi\alpha'_1)^{D/2}}E_N^{D/2-2}=
(D/2-2)\frac{\lambda_i^2}{(8\pi\alpha'_1)^{D/2}}E_N^{D_2-2}=(D/2-2)g_i^2.\]
So to find the anomalous dimensions we have only to multiply the renormalization constants by $(D/2-2)$.
Each of them contains $\Gamma(2-D/2)$. So we shall have a product
\[(D/2-2)\Gamma(2-D/2)=-\Gamma(3-D/2).\]

As a result
\beq
\gamma_1=\gamma_1^a+\gamma_1^b,\ \ \gamma_1^a=-\frac{1}{2}g_1^2\Gamma(3-D/2),
\ \ \gamma_1^b=\frac{1}{2}\frac{g_3^2}{u^{D/2}}\Gamma(3-D/2),\label{gamma1}
\eeq
\beq
\gamma_2=-\frac{g_2^2}{[(1+u)/2]^{D/2}}\Gamma(3-D/2).\label{gamma2}
\eeq

Finally we calculate $\tau_1$:
\[\tau_1=E_N\frac{\pd}{\pd E_N}\ln \Big(U_{1}^{-1}Z_1\Big)
=E_N\frac{\pd}{\pd E_N}\Big((Z_1-1)-(U_1-1)\Big).\]
Similarly, we can introduce
\[\tau_2=E_N\frac{\pd}{\pd E_N}\ln \Big(U_{2}^{-1}Z_2\Big)
=E_N\frac{\pd}{\pd E_N}\Big((Z_2-1)-(U_2-1)\Big).\]
From our expressions for $U_{i}$ we find
\[ E_N\frac{\pd}{\pd E_N}(U_{1}^a-1)=-\frac{1}{4}g_1^2\Gamma(3-D/2),\]
\[ E_N\frac{\pd}{\pd E_N}(U_{1}^b-1)=\frac{1}{4}\frac{g_3^2u}{u^{D/2}}\Gamma(3-D/2),\]
\[ E_N\frac{\pd}{\pd E_N}(U_{2}-1)=-\frac{g_2^2}{(1+u)[(1+u)/2]^{D/2}}\Gamma(3-D/2).\]
This gives
\beq
\tau_1=\tau_1^a+\tau_1^b,\ \ \tau_1^a=-\frac{1}{4}g_1^2\Gamma(3-D/2),
\ \ \tau_1^b=\frac{1}{4}\frac{g_3^2}{ u^{D/2}}(2-u)\Gamma(3-D/2),\label{tau1}\eeq
\beq\tau_2=-\frac{g_2^2}{[(1+u)/2]^{D/2}}\frac{u}{1+u}\Gamma(3-D/2).\label{tau2}\eeq

\subsection{Beta-functions}
To calculate $\beta$-functions one has to calculate the relevant diagrams for the non-trivial couplings.
In the single loop approximation which is our scope we have to calculate the adequate triangle diagrams.
This calculation is described in Appendix A. Here we present its results in the lowest order in small $\ep$.
The four $\beta$-function are
\beq
\beta_1=-\frac{1}{4}\epsilon g_1+\frac{3}{2}g_1^3- g_2g_3^2\frac{2}{u^2}+g_1g_3^2\frac{1+u}{4u^2},
\label{bet1}
\eeq
\beq
\beta_2
=-\frac{1}{4}\epsilon g_2+g_1g_2^2\frac{6+2u}{(1+u)^2}- g_2g_3^2\frac{1+8u-u^2}{4u^2(1+u)},
\label{bet2}
\eeq
\beq
\beta_3
=-\frac{1}{4}\epsilon g_3+g_1g_2g_3\frac{4}{1+u}+g_2^2g_3\frac{4}{u(1+u)^2}+ g_3^3\frac{u-1}{4u^2},
\label{bet3}
\eeq
and
\beq
\beta_4=g_1^2\frac{u}{4}-g_2^2\frac{4u^2}{(1+u)^3}+ g_3^2\frac{u-2}{4u}.
\label{bet4}
\eeq

\subsection{At $u\to 0$}

The region $u\to 0$ is clearly singular, since both the anomalous dimensions and $\beta$-functions
blow up as $u\to 0$ and finite $g_3$. Obviously, to make sense in this region also $g_3$ has to go to zero.
The physical meaning has the ratio $r=g_3/u$ in the limit $g_3,u\to 0$.
In terms of $r$ the anomalous dimensions associated with the pomeron are at $D=4$
\[\gamma_1^b=\frac{1}{2}r^2,\ \ \tau_1^b=\frac{1}{2}r^2,\]
so that $r$ has to be finite and of order $\sqrt{\ep}$ in the single loop approximation.
Therefore it is reasonable to pass to coupling constant $r$ instead of $g_3$ putting $g_3=ru$.
The new beta-function $\beta_r$, which governs the evolution of $r$, is easily found from $\beta_3$ and $\beta_4$:
\[\beta_r=E_N\frac{\partial r}{\partial E_N }=\frac{1}{u}(\beta_3-r\beta_4).\]
So expressing in coupling constants $g_1,\ g_2,\ r$ and $u$ we find new $\beta$-functions
\beq
\beta_1=-\frac{1}{4}\epsilon g_1+\frac{3}{2}g_1^3- 2g_2r^2+g_1r^2\frac{1+u}{4},
\label{bet1a}
\eeq
\beq
\beta_2
=-\frac{1}{4}\epsilon g_2+g_1g_2^2\frac{6+2u}{(1+u)^2}- g_2r^2\frac{1+8u-u^2}{4(1+u)},
\label{bet2a}
\eeq
\beq
\beta_r=r\Big(-\frac{1}{4}\ep-g_1^2\frac{1}{4} +g_1g_2\frac{4}{1+u}+4g_2^2\frac{1+u+u^2}{u(1+u)^3}+\frac{1}{4}r^2\Big)
\label{betha}
\eeq
and
\beq
\beta_4=g_1^2\frac{u}{4}-g_2^2\frac{4u^2}{(1+u)^3}+ r^2\frac{u(u-2)}{4}.
\label{bet4a}
\eeq

Inspection of $\beta_r$ shows that it is not finite when $u\to 0$ but blows up as $rg_2^2/u$.
Therefore for sensible evolution one has to impose conditions: either $g_2=0$ or $r=0$ or both $g_2=r=0$.
Remarkably, both conditions are compatible with evolution, since $\beta_2=0$ when $g_2=0$ and
$\beta_r=0$ when $r=0$. So if, for example, one chooses the initial $g_2=0$ then it will stay equal to zero during evolution
$\bar{g}_2(t)=g_2=0$. Similarly, if initially $r=0$ then $\bar{r}(t)=r=0$ at all $t$.

As a result, evolution in the vicinity of $u=0$ splits into three sectors depending on the initial conditions,
which are either $g_2=0$ or $r=0$ or $g_2=r=0$.

If the initial $g_2=0$ then $\bar{g}_2(t)=0$ and $g_1,\ r$ and $g_4$ evolve with $\beta$-functions
\beq
\beta_1=-\frac{1}{4}\epsilon g_1+\frac{3}{2}g_1^3+g_1r^2\frac{1+u}{4},
\label{bet1b}
\eeq
\beq
\beta_r=r\Big(-\frac{1}{4}\ep-g_1^2\frac{1}{4}+ \frac{1}{4}r^2\Big) ,
\label{bettb}
\eeq
\beq
\beta_4=g_1^2\frac{u}{4}+ r^2\frac{u(u-2)}{4}.
\label{bet4b}
\eeq
If the intial $r=0$ then $\bar{r}(t)=0$ and $g_1,\ g_2$ and $g_4$ evolve with $\beta$-functions
\beq
\beta_1=-\frac{1}{4}\epsilon g_1+\frac{3}{2}g_1^3,
\label{bet1c}
\eeq
\beq
\beta_2
=-\frac{1}{4}\epsilon g_2+g_1g_2^2\frac{6+2u}{(1+u)^2},
\label{bet2c}
\eeq
\beq
\beta_4=g_1^2\frac{u}{4}-g_2^2\frac{4u^2}{(1+u)^3}.
\label{bet4c}
\eeq
Finally, if both $g_2=0$ and $r=0$  then the remaining $g_1$ and $g_4$ evolve with $\beta$-functions
\beq
\beta_1=-\frac{1}{4}\epsilon g_1+\frac{3}{2}g_1^3
\label{bet1d}
\eeq
and
\beq
\beta_4=g_1^2\frac{u}{4}.
\label{bet4d}
\eeq

\section{At the fixed point}
\subsection{Scaling}

At the fixed point $g_i=g_{ic}$ we have
\beq\frac{d \bg_i(t)}{dt}=0,\ \  {\rm so\ that}\ \  \bg_i(t)=g_{ic}.
\label{barg}\eeq
Also,
\beq
\balpha'_1(-t)=\alpha'_1 e^{-tz},\ \ z=z(g_c)=1-\tau_1(g_c)\label{balpha}\eeq
and keeps  running.
The solution (\ref{eq056}) at $g=g_c$ becomes
\beq
\Gamma^R(\xi E_i, k_i,g_c,\alpha',E_N)=\Gamma^R(E_i,k_i,g_c,\alpha'_1e^{-zt},E_N)e^{t[1-
\sum_{i=1}^2(n_i+m_i)\gamma_i(g_c)/2]} .
\label{eq084}
\eeq

We use the scaling property
\[\Gamma^R(E_i,k_i,g_c,\alpha'_1,E_N)=E_N\Big(\frac{E_N}{\alpha'_1}\Big)^{(2-n-m)D/4}\Phi\Big(\frac{E_i}{E_N},
\frac{\alpha'_1}{E_N}{\bf k}_i{\bf k}_j,g_c\Big)\]
to obtain
\[\Gamma^R(\xi E_i,k_i,g_c,\alpha'_1,E_N)\]\[=e^{t[1-\sum_{i=1}^2(n_i+m_i)\gamma_{i}(g_c)/2]}E_N\Big(\frac{E_N}{\alpha'}\Big)^{(2-n-m)D/4}e^{tz(2-n-m)D/4}
\Phi\Big(\frac{E_i}{E_N},\frac{\alpha'_1 e^{-zt}}{E_N}{\bf k}_i{\bf k}_j,g_c\Big).\]
Rescaling here $E\to E/\xi$ we get
\[\Gamma^R(E_i,k_i,g_c,\alpha',E_N)=\]\[e^{t[1-\sum_{i=1}^2(n_i+m_i)\gamma_i(g_c)/2]}E_N\Big(\frac{E_N}{\alpha'}\Big)^{(2-n-m)D/4}e^{tz(2-n-m)D/4}
\Phi\Big(\frac{E_i}{E_N\xi},\frac{\alpha'_1 e^{-zt}}{E_N}{\bf k}_i{\bf k}_j,g_c\Big) .\]
Taking
\[\xi=\frac{-E}{E_N} ,\]
where $E$ is the sum of energies in the final state $n=\{n_1,n_2\}$,
we find finally
\[\Gamma^R(E_i,k_i,g_c,\alpha'_1,E_N)=E_N\Big(\frac{E_N}{\alpha'_1}\Big)^{(2-n-m)D/4}
\Big(\frac{-E}{E_N}\Big)^{1-\sum_{i=1}^2(n_i+m_i)\gamma_i(g_c)/2+z(2-n-m)D/4}\]\beq\times
\Phi\Big(\frac{-E_i}{E},\Big(\frac{-E}{E_N}\Big)^{-z}\frac{{\bf k}_i{\bf k}_j}{E_N}\alpha'_1,g_c\Big).\label{gamrphi}\eeq
In particular we find
\beq
\Gamma_i(E,k^2,g_c,\alpha'_1,E_N)=E_N\Big(\frac{-E}{E_N}\Big)^{1-\gamma_i(g_c)}\Phi_i(\rho),
\ \ i=1,2,
\label{eq087}
\eeq
where
\beq\rho=\Big(\frac{-E}{E_N}\Big)^{-z}\,\frac{\alpha'_1k^2}{E_N}.
\eeq
We see that $\Gamma_i$ has a zero at some point $\rho_{i0}(g_c)$ at which
\[\Phi_i(\rho_{i0},g_c)=0.\]
At this point
\[\Big(\frac{-E}{E_N}\Big)^{-z}\,\frac{\alpha'_1 k^2}{E_N}=\rho_{i0}(g_c),\ \ \
{\rm or}\ \
\frac{-E}{E_N}=\Big(\frac{\alpha'_1 k^2}{E_N\rho_{i0}(g_c)}\Big)^{1/z}.\]
This translates into the $(E,k^2)$- dependence
\[E=-E_N\Big(\frac{\alpha'_1 k^2}{E_N}\Big)^{1/z}f_i(g_c).\]
Since the zero of $\Gamma_i$ implies the singularity of the propagator,
the trajectory function $\alpha_i=1-E$ is found to be
\beq
\alpha_i(k^2)=1+E_N\Big(\frac{\alpha'_1 k^2}{E_N}\Big)^{1/z}f_i(g_c),\ \ i=1,2.
\label{eq089}
\eeq
Generally it  is not analytic at $k^2=0$.  In the model without odderon the slope is infinite at $k^2=0$  ~\cite{abarb1}.

\subsection{Scaling functions at the fixed point}

At the fixed point as $\epsilon\to 0$ constants $g_{1,2,3}^2$ and anomalous dimensions $\gamma_i$ and $(z-1)$ are proportional to $\epsilon$.
So the renormalized $\Gamma^R_i$ at $g=g_c$ are known in two first orders in the expansion in powers of $\epsilon$.
Comparing with its representation Eq. (\ref{eq087}) in terms of the scaling function $\Phi_j(\rho)$, $j=1,2$
we can find the scaling functions $\Phi_j$  in the two first orders in $\epsilon$.
Suppressing for the moment subindex $j=1,2$ in $\Phi_j$ we have in these orders
\[\Phi(\rho)=\Phi_{0}(\rho)+\epsilon\Phi_{1}(\rho)+...\ ,\]
where
\[\Phi_0(\rho)=\Phi_{\ep=0}(\rho).\]
Note that only the form of $\Phi$ is taken at $\ep=0$ but not the arguments, which are also $\ep$-dependent.

At the fixed point
\beq
\rho=\frac{\alpha'_1k^2}{E_N}e^{-tz(g_c)},\label{rho}\eeq
where
\[ t=L=\ln\frac{-E}{E_N}.\]

At $\ep=0$ we have
$z(0)=1$, so at zero order
\beq\rho_{\ep=0}=\rho_{0}=\frac{\alpha'_1 k^2}{-E}.
\label{rho0}
\eeq

In the first two orders in $\ep$ for both pomeron and odderon we get
\beq
\Gamma^R(E,k^2,g_c(\ep),\alpha'_1E_N)=-E
\Big\{\Phi_0(\rho_{i0})+
\ep\Big[\Phi_1(\rho_{0})-\gamma'(0)L\Phi_0(\rho_{0})
-z'(0)L\rho_{0}\frac{d\Phi_0(\rho_{0})}{\pd\rho_{0}}\Big]\Big\}.
\label{scfun1}
\eeq

We start from the zeroth order $\ep=0$. The inverse propagators are
for the pomeron
\[\Gamma^R_1=E-\alpha'_1k^2\]
and for the odderon
\[\Gamma^R_2=E-\alpha'_2k^2.\]
Separating factor $-E$
we find
\[\Gamma^R_1=-E(-1-\rho)\]
and
\[\Gamma^R_2=-E(-1-u\rho).\]
So in the lowest order
\beq
\Phi_{10}(\rho_0)=-1-\rho_0,
\label{phi10}
\eeq
\beq
\Phi_{20}(\rho_0)=-1-u\rho_0.
\label{phi20}
\eeq

In the linear order in $\ep$
\[\Gamma^R_j(E,k^2,g_i(\epsilon),\alpha'_1,E_N)_{linear\ in\ \ep}=
-\Sigma^R_j, \ \ j=1,2.\]
The renormalized self-mass for the pomeron is
\[\Sigma^R_1=\Sigma^R_a+\Sigma^R_b,\]
with
\[\Sigma^R_a=\epsilon d_1\Gamma(2-D/2)
\sigma_1\Big(\frac{(\sigma_1/E_N)^{D/2-2}}{1-D/2}+1\Big),\]
\[
\Sigma^R_b=\epsilon d_3\Gamma(2-D/2)
\sigma_3\Big(\frac{(\sigma_3/E_N)^{D/2-2}}{1-D/2}+1\Big).\]
For the odderon the renormalized self-mass is
\[
\Sigma^R_2=\epsilon d_2\Gamma(2-D/2)
\sigma_2\Big(\frac{(\sigma_2/E_N)^{D/2-2}}{1-D/2}+1\Big)
.\]
Here
\beq
\sigma_1=\frac{1}{2}\alpha'_{1}k^2-E,
\label{sms1}
\eeq
\beq
\sigma_2=\alpha'_1k^2\frac{u}{1+u)}-E,
\label{sms2}
\eeq
\beq\sigma_3=\frac{1}{2}\alpha'_2k^2-E.
\label{sms3}
\eeq
and
the constants $d_i$ are defined from
\[\epsilon d_1=-\gamma_1^a=\frac{1}{2}g_1^2,\ \ \epsilon d_2=-\gamma_2=\frac{g_2^2}{[(1+u)/2]^{D/2}},\ \
\epsilon d_3=-\gamma_1^b-\frac{g_3^2}{2u^{D/2}}.\]

At $\ep\to 0$ these expressions simplify.
We use
\[\Gamma(2-D/2)=\frac{2}{\ep},\ \ \frac{a^{D/2-2}}{1-D/2}=1+\frac{\ep}{2}(\ln a-1)\]
to get
\beq
[\Gamma(2-D/2)\frac{a^{D/2-2}}{1-D/2}+1=\ln a+1.
\label{eto0}
\eeq

Using (\ref{eto0}) we find in the limit $\ep\to 0$
\[\Sigma^R_a=\epsilon d_1\sigma_1
\Big(\ln\frac{\sigma_1}{E_N}-1\Big),\]
\[
\Sigma^R_b=\epsilon d_3
\sigma_3\Big(\ln\frac{\sigma_3}{E_N}-1\Big),\]
\[
\Sigma^R_2=\epsilon d_2
\sigma_2\Big(\ln\frac{\sigma_2}{E_N}-1\Big)
.\]

We express $\sigma_i$ via $\rho_{0}$ defining $\sigma_i=-E x_i$ with
\[x_1=1+\frac{1}{2}\rho_0,\]
\[x_2=1+\frac{u}{1+u}\rho_0,\]
\beq
x_3=1+\frac{1}{2}u\rho_{0}
\label{defxi}
\eeq
and rewrite the self-mass for the pomeron as
\[\Sigma^R_a=-\epsilon Ed_1x_1(L+\ln x_1-1),\]
\[
\Sigma^R_b=-\epsilon Ed_3
x_3(L+\ln x_3-1),\]
so that
\[\Sigma_1^R=-\ep E\Big(d_1x_1(\ln x_1-1)+d_3x_3(\ln x_3-1)\Big)-\ep EY_1 ,\]
where
\[ Y_1=L(d_1x_1+d_3x_3))=L\Big[d_1\Big(1+\frac{1}{2}\rho_0\Big)+d_3\Big(1+\frac{1}{2}u\rho_0)\Big)\Big]\]\beq
=L\Big[d_1+d_3+\frac{1}{2}\rho_0\Big(d_1+ud_3\Big)\Big].
\label{y1}
\eeq

For the odderon we get
\[\Sigma^R_2=-\epsilon Ed_2x_2(\ln x_2-1)-\ep EY_2 ,\]
where
\beq
Y_2=d_2L\Big(1+\frac{u}{1+u}\rho_0\Big).
\label{y2}
\eeq

In the linear order in $\ep$ we should have
\[
-\Sigma^R_i=-\ep E(\Phi_{i1}-X_i),\ \ i=1,2,\]
where
\beq
X_i=\gamma'_i(0)L\Phi_{i0}(\rho_{0})
+z'(0)L\rho_{0}\frac{d\Phi_{i0}(\rho_{0})}{\pd\rho_{0}}.
\label{xxi}
\eeq
So the scaling function linear in $\epsilon$ is given by
\beq
-\ep E\Phi_{i1}=-\ep EX_i-\Sigma^R_i.
\label{phii1}
\eeq

The coefficients in (\ref{xxi}) up to terms linear in $\ep$ are obtained as follows
\[\gamma_1=-\ep(d_1+d_3),\ \ \gamma_2=-\ep d_2,\]
\[\tau_1=-\ep\frac{1}{2}d_1+\ep\frac{1}{2}(2-u)d_3,\ \ z=1-\tau_1.\]

We start with $\Phi_{11}$.
We get
\[ \frac{X_1}{L}=\gamma'_1(0)\Phi_{10}+z'(0)\rho_0\frac{d\Phi_{10}}{d\rho}=
(-d_1-d_3)(-1-\rho_0)-\rho_0\frac{1}{2}\Big(d_1+(2-u)d_3\Big)\]\[=
d_1+d_3+\rho_0(d_1+d_3-d_1/2-(1-u/2)d_3)=d_1+d_3+\rho_0(d_1/2+ud_3/2).\]
One observes that
\[X_1=Y_1,\]
so that
\beq
\Phi_{11}(\rho_0)=-d_1x_1(\ln x_1-1)-d_3x_3(\ln x_3-1).
\label{phi11}
\eeq

Passing to the odderon we find
\[
\frac{X_2}{L}=\gamma'_2(0)\Phi_{20}+z'(0)\rho_0\frac{d\Phi_{20}}{d\rho}
=-d_2(-1-u\rho_0)-u\rho_0\frac{1}{2}\Big(d_1+(2-u)d_3\Big)\]\[=
d_2+\rho_0\Big(ud_2-\frac{1}{2}ud_1-\frac{u(2-u)}{2}d_3\Big).\]
The difference
\[X_2-Y_2=\rho_0\Big(ud_2=\frac{1}{2}ud_1-\frac{u(2-u)}{2}d_3-\frac{u}{1+u}d_2\Big)=
\rho_0\Big(\frac{1}{2}d_2-\frac{u^2}{1+u}d_2+\frac{u(2-u)}{2}d_3\Big),\]
or multiplying by $\ep$
\[\frac{\ep}{\rho_0}(X_2-Y_2)=
-\frac{u}{4}g_1^2+\frac{4u^2}{(1+u)^3}g_2^2-\frac{u(2-u)}{4u_2}g_3^2=-\beta_4,\]
which is zero at the fixed point.
So we get
\beq
\Phi_{21}(\rho_0)=-d_2x_2(\ln x_2-1).
\label{phi21}
\eeq

So collecting our results and using (\ref{eq087}) we get
\beq
\Gamma_1^R(E,k,g_c,\alpha'_1,E_N)=-E\Big(\frac{-E}{E_N}\Big)^{-\gamma_1(g_c)}
\Big[-1-\rho+\gamma_1^a(g_c)x_1(\ln x_1-1)+\gamma_1^b(g_c)x_3(\ln x_3-1)\Big)\Big]
\label{gamphi1}
\eeq
and
\beq
\Gamma_2^R(E,k,g_c,\alpha'_1,E_N)=-E\Big(\frac{-E}{E_N}\Big)^{-\gamma_2(g_c)}
\Big[-1-u\rho+\gamma_2(g_c)x_2(\ln x_2-1)\Big],
\label{gamphi2}
\eeq
where now
\[x_1=1+\frac{1}{2}\rho,\]
\[x_2=1+\frac{u}{1+u}\rho,\]
\beq
x_3=1+\frac{1}{2}u\rho
\eeq
and $\rho$ is given by (\ref{rho}).

At $k=0$ we have $\rho=0$ and $x_1=x_2=x_3=1$, so that taking into account that
$\ep d_2=-\gamma_2$ and $\ep(d_1+d_3)=-\gamma_1$ we obtain
\beq
\Gamma_1^R(E,k=0,g_c,\alpha'_1,E_N)=E\Big(\frac{-E}{E_N}\Big)^{-\gamma_1(g_c)}
(1+\gamma_1)
\label{gamphi10}
\eeq
and
\beq
\Gamma_2^R(E,k=0,g_c,\alpha'_1,E_N)=E\Big(\frac{-E}{E_N}\Big)^{-\gamma_2(g_c)}
(1+\gamma_2).
\label{gamphi20}
\eeq

To know the behaviour of the propagators at high energies we have to perform the inverse Laplace transformation, which requires to
calculation of
\[ I(a)=\frac{1}{2\pi i}\int_{c-i\infty}^{c+\infty}dE\ e^{-Ey} E^{-1+a},\]
where the contour should be taken to the left of the singular point $E=0$. Note that this position is
opposite to the normal one, since $E=1-\alpha(t)$.
The integral reduces to the one over the upper side of the right cut multiplied by $1-e^{i\pi a}$
\[I(a)=-\frac{1-e^{i \pi a}}{2\pi i}\int_0^\infty dE\ e^{-yE}E^{-1+a}
=-\frac{1-e^{i\pi a}}{2\pi i}y^{-a}\Gamma(a).\]
At small $a$ the right-hand side is
\[I(a)=-\frac{1-e^{i\pi a}}{2\pi i}y^{-a}\Gamma(a)y^{-a}\simeq y^{-a}a\Gamma(a)=y^{-a}.\]
So as a function of the physical c.m. energy squared $s=e^y$ the pomeron and odderon propagators at $k=0$ are
\beq
P_{pomeron}(s,k^2=0,g_c,\alpha'_1,E_N)=\Big(1+\gamma_1(g_c)\Big)s(\ln s)^{-\gamma_1(g_c)},
\label{ppom}
\eeq
\beq
P_{odderon}(s,k^2=0,g_c,\alpha'_1,E_N)=\Big(1+\gamma_2(g_c)\Big)s(\ln s)^{-\gamma_2(g_c)}.
\label{podd}
\eeq

\section{Real fixed points}

As is well known without odderon in the pomeron model with a single coupling constant one has a single
attractive fixed point  $g_c=\sqrt{\ep/6}$.
Inclusion of the odderon complicates the situation. First, the number of non-trivial fixed point is raised to five.
Second, attraction or repulsion in the 4-dimensional space of the four coupling constants may be
attractive along certain directions in this space and repulsive along the rest ones. Depending on their numbers
we can have different grades of attraction and repulsion from zero to four.
We postpone a  more detailed discussion of this point until the next section.
The found five real fixed points at which all $\beta$-functions vanish  will be denoted by a set of coupling constants
\[g_c=\{g_1,g_2,g_3,g_4\}.\]
Additionally at $g_3=g_4=0$ we show the value of the ratio $r=g_3/g_4$, which characterizes the
anomalous dimensions associated with the odderon.
Derivation of some less trivial fixed points is described in Appendix B.

\subsection{Fixed point $g_c^{(0)}$ with $g_1=g_2=g_4=0$}

At this fixed point the only interaction is transition of the pomeron into a pair of odderons:
\[ g_c^{(0)}=\{0,\ 0,\ 0,\ 0\},\ \ r=g_3/g_4=\sqrt{2}\sqrt{\ep/2} .\]
The anomalous dimensions are
\[\gamma_1^a=0,\ \ \gamma_1^b=\frac{\ep}{2},\ \ \gamma_2=0,\ \ \tau_1=\frac{\ep}{2}.\]
The scaling functions are
\[\Phi_1(\rho)=-1-\rho-\frac{\ep}{2},\]
\[\Phi_2=-1.\]

So
\beq
\Gamma^R_1(E,k^2=0,g_1,\alpha'_1,E_N)=\Big(1+\frac{\epsilon}{2}\Big)E\Big(\frac{-E}{E_N}\Big)^{-\epsilon/2},
\label{gampom0}
\eeq
\beq
\Gamma^R_2(E,k^2=0,g_1,\alpha'_1,E_N)=E.
\label{gamod0}
\eeq
As a function of energy squared $s$ the pomeron and odderon propagators are
\beq
P_{pomeron}(s,k^2=0,g_c,\alpha'_1,E_N)=\Big(1+\frac{\epsilon}{2}\Big)s(\ln s)^{-\epsilon/2},
\label{ppom0}
\eeq
\beq
P_{odderon}(s,k^2=0,g_c,\alpha'_1,E_N)=s .
\label{podd0}
\eeq
It corresponds to the cross-section slowly (logarithmically) going down at high energies.
The negative signature amplitude rises as $s$ but does not contribute to the cross-section, since its
contribution to the scattering amplitude is real. The latter behaviour is also found at the following
fixed points $g_p^{(1)}$ and $g_p^{(4)}$ at which $g_2=0$.

\subsection{Fixed point $g_c^{(1)}$ with $g_2=g_3=g_4=0$ and $r=g_3/g_4=0$}

This is the same   fixed point  as without odderon.
\[g_c^{(1)}=\{\frac{\sqrt{\epsilon/2}}{\sqrt{3}},0,0,0\},\ \ r=0.\]
It does not include the odderon at all whose behaviour remains purely perturbative. The pomeron sector
exactly corresponds to the case discussed in ~\cite{abarb2,abarb1}.
The anomalous dimensions  are
\[\gamma_1^a=-\frac{\ep}{12},\ \ \gamma_1^b=0,\ \ \gamma_2=0,\ \ \tau_1=-\frac{\ep}{24}.\]
The scaling functions are
\[\Phi_1(\rho)=-1-\rho-\frac{\epsilon}{12}(1+\rho/2)[\ln (1+\rho/2)-1],\]
\[\Phi_2=-1 .\]

At $k=0$
\[\Phi_1(0)=-1+\frac{\epsilon}{12},\ \ \Phi_2(0)=-1\]
and so
\beq
\Gamma^R_1(E,k^2=0,g_1,\alpha'_1,E_N)=\Big(1-\frac{\epsilon}{12}\Big)E\Big(\frac{-E}{E_N}\Big)^{\epsilon/12},
\label{gampom1}
\eeq
\beq
\Gamma^R_2(E,k^2=0,g_1,\alpha'_1,E_N)=E.
\label{gamod1}
\eeq
As a function of energy squared $s$ the pomeron and odderon propagators are
\beq
P_{pomeron}(s,k^2=0,g_c,\alpha'_1,E_N)=\Big(1-\frac{\epsilon}{12}\Big)s(\ln s)^{\epsilon/12}.
\label{ppom1}
\eeq
\beq
P_{odderon}(s,k^2=0,g_c,\alpha'_1,E_N)=s.
\label{podd1}
\eeq

\subsection{Fixed point $g_c^{(2)}$ with $g_3=0$ and $g_4\neq 0$}

This fixed point $\tg_c$ is given by the set
\[g_{1}=\frac{\sqrt{\epsilon}}{\sqrt{6}}=0.5773\sqrt{\epsilon/2},\ \
g_{2}=\frac{\sqrt{3}(1+u_c)^2}{12+4u_c}\sqrt{\epsilon/2}= 0.3975\sqrt{\epsilon/2},\]
\[g_{4}=u_c=\frac{3}{16}(\sqrt{33}-1)=0.8896\]
and $\tg_3=0$.

In this case
\[\gamma_1^a=-\frac{\ep}{12},\ \ \gamma_1^b=0,\ \ \gamma_2=-\frac{\ep}{11.30},\ \ \tau_1=-\frac{\ep}{24}, \]
so the final scaling functions are
\beq
\Phi_1(\rho)=-1-\rho-\frac{\epsilon}{12}\Big(\frac{1}{2}\rho+1\Big)\Big[\ln\Big(\frac{1}{2}\rho+1\Big)-1\Big],
\eeq
\beq
\Phi_2(\rho)=-1-u\rho-\frac{\epsilon}{11.30}\Big(\frac{u}{1+u}\rho+1\Big)\Big[\ln\Big(\frac{u}{1+u}\rho+1\Big)-1\Big],
\eeq
where we recall that
\[u=u_c=0.8876.\]

When $k^2=0$
\[\Phi_1(0)=-1+\frac{\epsilon}{12},\ \ \Phi_2(0)=-1+\frac{\epsilon}{11.30}\]
and we find
\beq
\Gamma^R_1(E,k^2=0,g_1,\alpha'_1,E_N)=\Big(1-\frac{\epsilon}{12}\Big)E\Big(\frac{-E}{E_N}\Big)^{\epsilon/12},
\label{gampom2}
\eeq
\beq
\Gamma^R_2(E,k^2=0,g_1,\alpha'_1,E_N)=\Big(1-\frac{\epsilon}{11.30}\Big)E\Big(\frac{-E}{E_N}\Big)^{\epsilon/11.30}.
\label{gamod2}
\eeq

As a function of energy squared $s$ the pomeron and odderon propagators are
\beq
P_{pomeron}(s,k^2=0,g_c,\alpha'_1,E_N)=\Big(1-\frac{\epsilon}{12}\Big)s(\ln s)^{\epsilon/12},
\label{ppom2}
\eeq
\beq
P_{odderon}(s,k^2=0,g_c,\alpha'_1,E_N)=\Big(1-\frac{\epsilon}{11.30}\Big)s(\ln s)^{\epsilon/11.30}.
\label{podd2}
\eeq

\subsection{Fixed point $g_c^{(3)}$ with $g_3=g_4=0$ and $r=g_3/g_4=0$}

This is the third fixed point which appears when  $u=0$ but $g_2\neq 0$
\[ g_c^{(3)}=\{\frac{\sqrt{\ep/2}}{\sqrt{3}},\ \frac{\sqrt{\ep/2}}{\sqrt{48}},\ 0,\ 0\},\ \  r=g_3/g_4=0.\]

In this case
\[\gamma_1^a=-\frac{\ep}{12},\ \ \gamma_1^b=0,\ \ \gamma_2=-\frac{\ep}{24},\ \ \tau_1=-\frac{\ep}{24} .\]

Since $u=0$ the scaling function for the odderon simplifies:
\[\frac{u}{1+u}\rho+1\to 1.\]
So the final scaling functions are
\beq
\Phi_1(\rho)=-1-\rho-\frac{\epsilon}{12}\Big(\frac{1}{2}\rho+1\Big)\Big[\ln\Big(\frac{1}{2}\rho+1\Big)-1\Big],
\eeq
\beq
\Phi_2(\rho)=-1+\frac{\epsilon}{24}.
\eeq

When $k^2=0$
\[\Phi_1(0)=-1+\frac{\epsilon}{12},\ \ \Phi_2(0)=-1+\frac{\epsilon}{24}\]
and we find
\beq
\Gamma^R_1(E,k^2=0,g_1,\alpha'_1,E_N)=\Big(1-\frac{\epsilon}{12}\Big)E\Big(\frac{-E}{E_N}\Big)^{\epsilon/12},
\label{gampom3}
\eeq
\beq
\Gamma^R_2(E,k^2=0,g_1,\alpha'_1,E_N)=\Big(1-\frac{\epsilon}{24}\Big)E\Big(\frac{-E}{E_N}\Big)^{\epsilon/24}.
\label{gamod3}
\eeq
As a function of energy
\beq
P_{pomeron}(s,k^2=0,g_c,\alpha'_1,E_N)=\Big(1-\frac{\epsilon}{12}\Big)s(\ln s)^{\epsilon/12},
\label{ppom3}
\eeq
\beq
P_{odderon}(s,k^2=0,g_c,\alpha'_1,E_N)=\Big(1-\frac{\epsilon}{96}\Big)s(\ln s)^{\epsilon/24}.
\label{podd3}
\eeq

\subsection{Fixed point $g_c^{(4)}$ with $g_1=g_2=0$}

In this case we have the fixed point
\[g_c^{(4)}=\{0,\ 0,\ 2\sqrt{2} \sqrt\frac{\epsilon}{2},\ 2\}.\]
Anomalous dimensions are
\[\gamma_1^a=0,\ \ \gamma_1^b=\frac{\ep}{2},\ \gamma_2=0,\ \ \tau_1=0 .\]

As a result we find the scaling functions
\[\Phi_1(\rho)=-1-\rho+\frac{\epsilon}{2}(\rho+1)\Big[(\ln (\rho+1)-1\Big],\]
\[\Phi_2(\rho)=-1-2\rho.\]

When $k^2=0$ we find
\[\Phi_1(0)=-1-\frac{\epsilon}{2},\ \ \Phi_2(0)=-1\]
and
\beq
\Gamma^R_1(E,k^2=0,g_1,\alpha'_1,E_N)=\Big(1+\frac{\epsilon}{2}\Big)E\Big(\frac{-E}{E_N}\Big)^{-\epsilon/2},
\label{gampom4}
\eeq
\beq
\Gamma^R_2(E,k^2=0,g_1,\alpha'_1,E_N)=E.
\label{gamod4}
\eeq
As a function of energy
\beq
P_{pomeron}(s,k^2=0,g_c,\alpha'_1,E_N)=\Big(1+\frac{\epsilon}{2}\Big)s(\ln s)^{-\epsilon/2},
\label{ppom4}
\eeq
\beq
P_{odderon}(s,k^2=0,g_c,\alpha'_1,E_N)=s.
\label{podd4}
\eeq
The behaviour at large energies is the same as at $g_p^{(0)}$:
it corresponds to the cross-section slowly (logarithmically)
going down at high energies.

One can note that the asymptotical cross-section (at $\epsilon=2$)
determined by the fixed points $g_c^{(1)}$, $g_c^{(2)}$ and $g_c^{(3)}$
grows as $(\ln s)^{1/6}$, what coincides with the result for the model
without odderon~\cite{abarb1}, but for the cases of fixed points
$g_c^{(0)}$ and $g_c^{(4)}$ it decreases as $(\ln s)^{-1}$.

\section{Close to the fixed point}

\subsection{Attractive domains}

The movement of the coupling constants $\bg(t)$ in the vicinity of the fixed point is determined by the matrix of
derivatives $\{\pd\beta_i/\pd g_k\}$ calculated at the fixed point.
For technical reasons we use the twice bigger matrix $a=\{a_{ik}\}=\{2\pd\beta_i/\pd g_k\}$  (see Appendix A).
Accordingly for evolution we use twice smaller parameter  $\tau=-t/2$.
Let eigenvalues of matrix $a$ be $x=\{x^{(1)},\ x^{(2)}, \ x^{(3)},\ x^{(4)}\}$ and the corresponding eigenvectors be $v^{(j)}$, $j=1,...,4$
where $v^{(j)}=\{v^{(j)}_1,\ v^{(j)}_2,\ v^{(j)}_3,\ v^{(j)}_4\}$. One can develop $\bg(t)$, $g$
and $g_c$ in these
eigenvectors:
\[\bg(t)=\sum_{j=1}^4\balpha^{(j)}(t)v^{(j)},\ \ \balpha^{(j)}(t)=\frac{<\bg(t)|w^{(j)}>}{N^{(j)}},\]
\beq
g=\sum_{j=1}^4\alpha^{(j)}v^{(j)},\ \ \alpha^{(j)}=\frac{<g|w^{(j)}>}{N^{(j)}},\ \
g_c=\sum_{j=1}^4\alpha_c^{(j)}v^{(j)},\ \ \alpha_c^{(j)}=\frac{<g_c|w^{(j)}>}{N^{(j)}},
\label{devtkop}\eeq
where $w^{(j)}$ are the eigenvectors of the transposed matrix $a^T$, orthonormalized according to
\beq
<w^{(j)}|v^{(l)}>=N^{(j)}\delta_{jl},\ \  j,l=1,...,4.\label{norm}\eeq
Then the movement of $\bg(t)$ will be given by
\beq
\bg(t)=g_c+\sum_{j=1}^4e^{-t x^{(j)}}(\alpha^{(j)}-\alpha_c^{(j)}).
\label{bgt}
\eeq
Note that in our calculations we have to take $\bg(-t)$ and afterwards take $t\to -\infty$,
so that in the end in (\ref{bgt}) $-t=2\tau\to\infty$.

We see that the movement of $\bg(t)$ depends on the signs of the real parts of eigenvalues $x^{(j)}$,
Along the direction of $x^{(j)}$ with positive real parts the fixed point is attractive and along the directions
of $x^{(j)}$ with  negative real parts the fixed point is repulsive.
To exclude running away from the fixed point one has to impose  conditions on the initial $g$
\[\alpha^{(j)}-\alpha_c^{(j)}=0,\ \ {\rm for\ Re}\, x^{(j)}<0,\]
or
\beq
<g|w^{(j)}>=<g_c|w^{(j)}>,\ \ {\rm for\  Re}\,x^{(j)}<0.
\label{attdom}
\eeq
These conditions define the domain of initial points  in which they are attracted to the fixed point,
''attractive domain''. Its dimension is given by the number of
 eigenvalues with positive real parts.

Matrix $a$ depends on $\epsilon$ non-trivially.
 In the following we consider the physical case
 $\epsilon=2$.

\subsection{Fixed point $g_c^{(0)}$ with only $r\neq 0$}

In this case $g_2=0$ and evolution is governed by the 3$\times$3 matrix
derived from (\ref{bet1b}) -- (\ref{bet4b}).
Matrix $a$ at the fixed point is diagonal with three diagonal matrix elements $x_1,\ x_r$ and $x_4$
 \[ x=\{ 0,\ 2,\ -2\}.\]
 Constant $g_1$ is not moving but to stay at the fixed point it has to be taken zero.
 So the attractive domain is 1-dimensional and determined by equations
 \[g_1=g_2=u=0.\]
 Unless the initial conditions lie on this line evolution will drive the coupling constants
 $g_2$ and $g_4$ away from the fixed point.

\subsection{Fixed point $g_c^{(1)}$ with only $g_1\neq 0$}

 In this  case both $g_2=0$ and $r=0$ so that evolution is  governed by the  2$\times$2 matrix
derived from (\ref{bet1d}),(\ref{bet4d}).
 The corresponding 2$\times$2 matrix
 $a$ is also diagonal with matrix elements $x_1$ and $x_4$
 \[ x=\{ 2,\ 1/6\}.\]
 Both $g_2$ and  $r$ are not moving but to stay at the fixed point they have to be taken zero.
 So the fixed point is purely attractive in the 2-dimensional domain
 determined by equations
 \[g_2=r=0\ \ {\rm or}\ \  g_2=g_3=0.\]
 Unless the initial conditions lie on this surface evolution will drive the coupling constants
 $g_2$ and $g_3$ away from the fixed point.
 This is  confirmed by numerical calculations in Section 7.

\subsection{Fixed point $g_c^{(2)}$ with $g_3=0$}

Calculations  give the following numerical results.\\
Eigenvalues $x=(x^{(1)},x^{(2)},x^{(3)},x^{(4)})$ of matrix $a$  are
\[ x= (2.0000,\  1.2085,\ 0.36956,\ -0.13976). \]
So the first three eigenvalues are positive, but the fourth $x^{(4)}$ is negative.

The corresponding eigenvectors $v^{(j)}$ of matrix $a$ and $w^{(j)}$ of matrix $a^T$ form matrices $v=\{v_{j.i}\}$
and $w=\{w_{ji}\}$:
 \[v=\left(
 \begin{array}{cccc}
 -0.75857& -0.52227&  0& 0\\
  0& -0.74105&  0&  0.4850\\
    0&  0&  1&  0\\
  0&  0.42028&  0&  1.5038
  \end{array}\right),\ \
w=\left(
\begin{array}{cccc}
  1&  0&  0&  0\\
 -0.66308&  0.96310&  0& -0.26916\\
  0&  0 & 1&0\\
 -0.39555&  0.57452&  0&  0.8777

\end{array}\right).
\]

Using the found eigenvectors it is trivial to solve the evolution equation (\ref{bgt}).
 Below we report the results of numerical calculations along these formulas.
With $g_3=0$ the fixed point $\tg_c=(\tg_{1c},\tg_{2c},\tg_{3c},\tg_{4c})$ is
\[
\tg_c= (0.577350,\  0.397500,\  0,\  0.889605),\ \ \tep=2.\]

\begin{figure}[h!p]
\begin{center}
\includegraphics[width=8 cm]{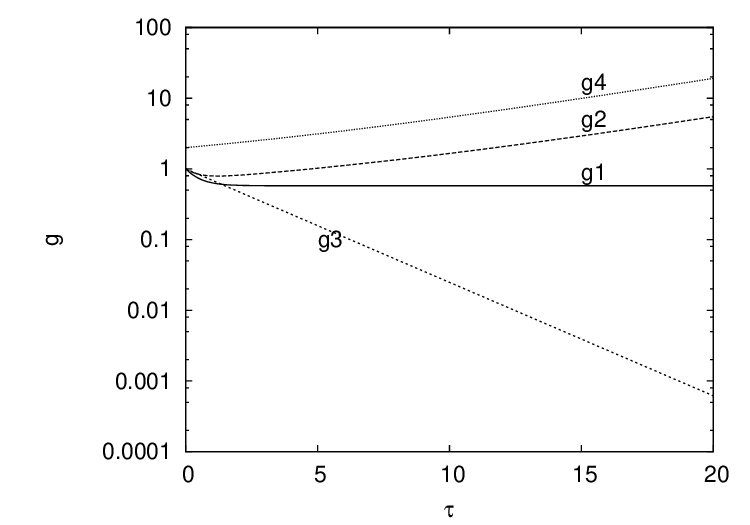}
\includegraphics[width=8 cm]{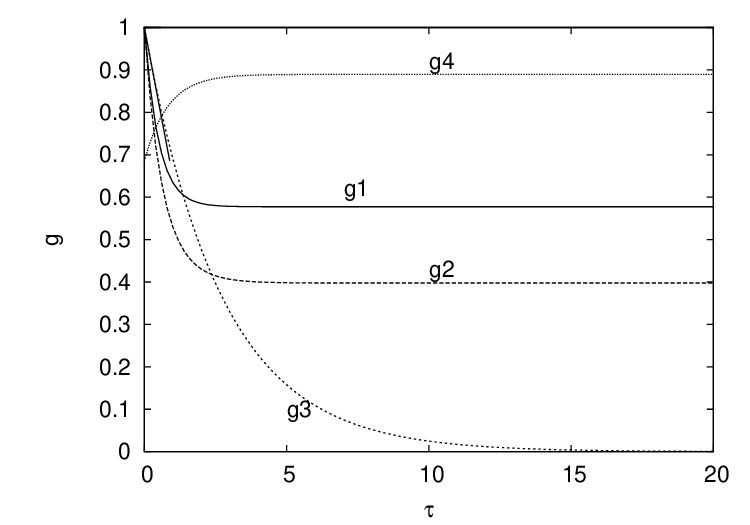}
\caption{Fixed point with $g_3=0$. Evolution of the coupling constants in the vicinity of the fixed point as $\tau=- t/2$ grows
with the initial values (\ref{ini}) outside the attractive domain (left panel) and with initial values (\ref{ini1})
inside this domain (right panel).}
\label{fig2}
\end{center}
\end{figure}

 To study evolution we have to start with some initial conditions. Rather arbitrarily we choose
 $\tg=(\tg_1,\tg_2,\tg_3,\tg_4)$ as
 \beq
 \tg=(  1,\  1,\  1,\  2),\ \ \tep=2.
 \label{ini}
 \eeq
 Using development in eigenvectors we find $g(\tau)$ shown in Fig.~\ref{fig2} in the left panel.
As one observes, coupling constants $\tg_1$ and $\tg_3$ flow to their respective values at the fixed point.
However, coupling constants $g_2$ and $g_4$ move away from the fixed point and grow. This exhibits the structure
of the fixed point considered: both $\tg_1$ and $\tg_3$ evolve independently of each other and from $\tg_2$ and $\tg_4$,
which mix and include the repulsive direction.

To find the subspace of coupling constants converging to the fixed point we have to exclude the growing
contribution using restriction (\ref{attdom}). Solving it for $\tg_4$ we have to set
\beq
\tg_4=\frac{1}{w^{(4)}_4}\Big(<\tg_c|w>-\sum_{i=1}^3 \tg_iw^{(4)}_i\Big).
\label{adjust}
\eeq
With $\tg_{1,2,3}$ given by (\ref{ini}) we get $\tg_4=0.6857$ so that the new initial values are
\beq
 \tg=(  1,\  1,\  1,\ 0.6857).
 \label{ini1}
 \eeq
Evolution with these new initial values is shown in Fig.~\ref{fig2} in the right panel.
One observes indeed that with $g$ given by (\ref{ini1}) all four coupling constants tend to their values at fixed point as
$|t|\to\infty$.

\subsection{Fixed point $g_c^{(3)}$ with $g_3=g_4=0$ and $r=g_3/g_4=0$}

In this case at the fixed point we have $g_3=g_4=0$ with $g_3/g_4=0$
and evolution of $g_1,g_2$ and $g_4$ is governed by the 3$\times$3 matrix
derived from
(\ref{bet1c}), (\ref{bet2c}), (\ref{bet4c}).
Calculated at the fixed point it is diagonal with all three eigenvalues $x_1,\ x_2$ and $x_4$ positive
\[ x=\{2,\ 1,\ 1/6\}.\]
So the only remaining condition for the attractive domain is
\[r=g_3/g_4=0.\]

It is instructive to see that how the same result can be numerically
obtained  using the full 4-dimensional space of four coupling constants $g_i$ and
 $\beta$-functions $\beta_i$, $i=1,...,4$.
The stability matrix $a$ then actually does not exist at $u=0$. Indeed in this case
\[a_{33}= -1+8g_1g_2+8g_2^2\frac{1}{u}\simeq \frac{1}{6u}\]
and so has a pole at $u=0$
Accordingly at small $u$ the eigenvalues of $a$  are found to be
\[ x=\{2,\ 1,\ \frac{1}{6},\ \frac{1}{6u}\}.\]
So the fourth eigenvalue goes to infinity at $u\to 0$ but remarkably staying positive,
since the case $u\to 0$ actually corresponds to $\alpha'_2\to 0$ but staying positive on
physical grounds.
Evolution of eigenvectors contains factors $\exp(-tx^{(j)}/2)$ which have finite limits when
$x^{(j)}\to +\infty$. So we expect to obtain reasonable results taking sufficiently small but finite values of $g_{4}$.
In the following calculation we take $u=10^{-5}$. Taking $u$ slightly above zero, but $g_3=0$ we mimic the
condition $r=0$ implied in the existence of this fixed point  and exploited in the above 3-dimensional treatment.
With all $x^{(j)}$ , $j=1,...,4$ positive the fixed point $g_c^{(3)}$ is  fully attractive
 although with the marked singular behaviour in the fourth direction.

The corresponding eigenvectors $v^{(j)}$ of matrix $a$ and $w^{(j)}$ of matrix $a^T$ form matrices $v=\{v_{j.i}\}$
and $w=\{w_{ji}\}$:
 \[v=\left(
\begin{array}{llll}
  1&  0.25&  0&  0.3149\cdot 10^{-5}\\
  0& -0.3608\cdot 10^{10}&    0&  1\\
  0& 0&  1&  0\\
  0&  0.2886&  0
\end{array}\right),
\]\[
w=\left(
\begin{array}{llll}
  1&  0&  0&  0\\
  0.8659& -3.464&  0&  1\\
  0&  0&  1&  0\\
 -0.3149\cdot 10^{-5}&  0.2771\cdot 10^{-9}&  0&  1
 \end{array}\right).
 \]

\begin{figure}[!h]
\begin{center}
\includegraphics[width=8 cm]{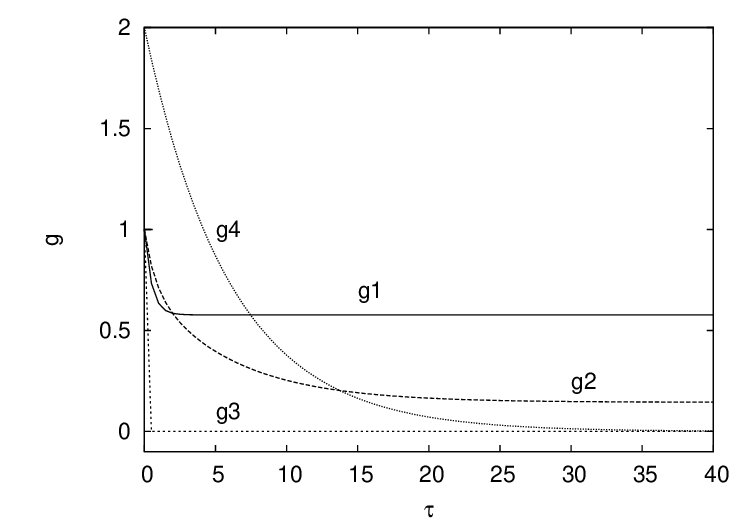}
\caption{Fixed point with $g_3=g_4=0$. Evolution of the coupling constants in the vicinity of the fixed point as $\tau=- t/2$ grows
with the initial values (\ref{ini}).}
\label{fig3}
\end{center}
\end{figure}

Using the found eigenvectors we solve the evolution equation (\ref{bgt}).
Below we report the results of numerical calculations along these formulas.
 For the initial conditions we take the same (\ref{ini}):
 \[
 \tg=(  1,\  1,\  1,\  2),\ \ \tep=2.
 \]
 Using the development in eigenvectors we find $g(\tau)$ shown in Fig.~\ref{fig3}.
This fixed point is attractive in all directions. So, as expected, the coupling constants tend to their values at the fixed point.
In absence of negative eigenvalues $x^{(j)}$ no adjustment is needed. As mentioned, condition $g_3/g_4=0$ at small $g_4$
is automatically fulfilled in the calculation. The astonishing fact is a very slow going to the
limiting values $g_c$ as compared to the previous case with adjusted initial coupling constant when the limits were
practically reached already at $|t|\sim 5$. Now these limits are only reached for $g_2$ and especially $g_4$ at $|t|$ of the order 100.
Note also the (also expected) practically immediate  drop of $g_3$ from  its initial value $g_3=1$ to its final value $g_3=0$
at the fixed point. Obviously with $u$ taken exactly zero this drop will instantly occur at $t=0$.

\subsection{Fixed point $g_c^{(4)}$ with $g_1=g_2=0$}

The eigenvalues of the stability matrix  $a$ are found to be
\beq
x=\{2,\ -\frac{16}{3},\ 2,\ 2\}.
\label{eig121}
\eeq
The four eigenvectors $v_i^{(j)}\equiv\{v_{ij}\}$ of matrix $a$
and the four eigenvectors  $w_i^{(j)}\equiv\{w_{ji}\}$ of matrix $a^T$ for eigenvalues $x^{(j)}$
given by (\ref{eig121}) are
\[v=\left(
\begin{array}{cccc}
1&0&0&0\\
12/11&1&0&0\\
0&0&1&0\\
0&0&0&1
\end{array}
\right),\ \ \
w=\left(
\begin{array}{cccc}
1&-12/11&0&0\\
0&1&0&0\\
0&0&1&0\\
0&0&0&1
\end{array}\right).\]

\begin{figure}[!h]
\begin{center}
\includegraphics[width=8 cm]{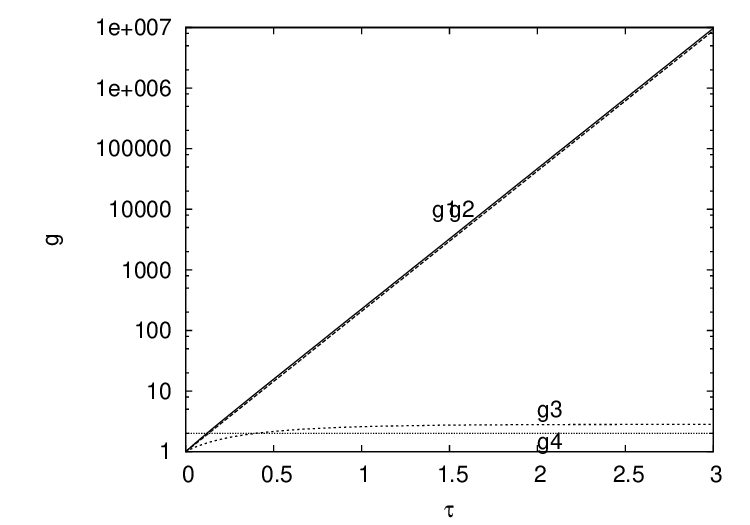}
\includegraphics[width=8 cm]{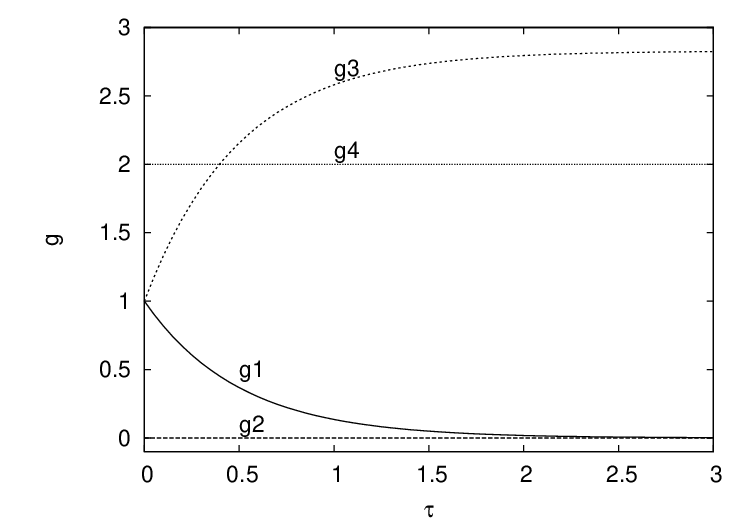}
\caption{Fixed point with $g_1=g_2=0$. Evolution of the coupling constants in the vicinity of the fixed point as $\tau=- t/2$ grows
with the initial values (\ref{ini}) outside the attractive domain (left panel) and with initial values $g=\{1,\ 0,\ 1,\ 2\}$
inside this domain (right panel).}
\label{fig4}
\end{center}
\end{figure}

The flow in the vicinity of the considered fixed point is clear from the  simple structure of eigenvalues and eigenvectors.
If the initial point $g$ is chosen arbitrarily then its behaviour with $t$ will be determined by the
contribution with $v^{(2)}$ and with the growing $t$ the point will go away from the fixed  point as $\exp(16 \tau/3)$
(we define $\tau=- t/2$). To find a subspace in which this growth does not take place and the point is attracted to the fixed point,
as discussed above, we have to impose a restriction $<g|w^{(2)}>=<g_c|w^{(2)}>$. With the form of $g_c$, $g$ and $w^{(2)}$ this restriction
reduces to condition
\[ g_2=0.\]
In the 4-dimensional space of all initial $g$ this condition cuts a 3-dimensional subspace of points which go to the fixed point at large $t$.
This is illustrated in Fig.~\ref{fig4} where we compare evolution from our previous initial values (\ref{ini}) $g=\{1,\ 1,\ 1,\ 2\}$
shown in the left panel and changed initial values $g=\{1,\ 0.\ 1,\ 2\}$ which satify condition (\ref{attdom}) in the right panel.

\section{Away from the fixed point}

Evolution of the coupling constants with growing $|t|$ is determined by equations
\beq
\frac{d\bg_i(t)}{dt}=-\beta_i(\bg(t)),\ \ i=1,...,4,
\label{evol}
\eeq
where $\beta$ -functions are presented in Section 3 and at $t=0$ the initally conditions
\[\bg_i(0)=g_i,\ \ i=1,...,4\]
are given.
In the previous section we studied this evolution in an approximate manner
linearizing $\beta$-functions in the vicinity of a chosen fixed point. Such evolution
can only take place while the coupling constants stay close to the chosen fixed point.
Taking an arbitrary initial coupling constants $g$ one can study evolution using  Eqs.
(\ref{evol}) with full $\beta$-functions, which depend on $\bg$ nonlinearly.
Such full evolution can easily be made numerically for any given initial values.

We studied evolution from the hypercube of initial coupling  constants
\beq
g_i\in [0,\ 1],\ \ i=1,...,4
\label{inidom}
\eeq
divided in 20 subintervals with $\Delta g=0.05$. To avoid divergence we took $g_4\geq 0.05$,
so that at $g_3=0$ the ratio $r=g_3/g_4=0$, which prohibits attraction to $g_c^{(0)}$.
So excluding cases with $g_{1,2,3}=0$, trivial with this restriction,
we obtain 185200 trajectories starting from this domain towards higher $|t|$.
Since we obviously cannot show them all graphically, we just describe the results in words.

First, nearly one third (65622 cases) of the trajectories go to infinity. Especially fast go to infinity
trajectories starting from $g_1=g_3=0$ with arbitrary $g_2$ and $g_4$.
In them $g_4$ attains values of the order $10^{13}$ at $\tau=30$. In the rest of trajectories going to infinity
this order of $g_4$ is only attained at $\tau$ of the order 100.

Second, all the trajectories which do not go to infinity (119578 cases) tend to one of the three fixed points $g_c^{(1)}$, $g_c^{(3)}$
and $g_c^{(4)}$. The fixed point $g_c^{(2)}$ does not appear as a result of evolution. The three mentioned fixed
points are only reached at very high values of $\tau$ of the order 100.

The division of final fixed points reached in 119578 cases is
\[g_c^{(1)}:g_c^{(2)}:g_c^{(3)}:g_c^{(4)}=400:0:110778:8400.\]
This division can be explained if we assume that the main property
of the stability domain, namely, the number of attractive
and repulsive directions, is retained while away from the  relevant
fixed points. For this reason with the arbitrary chosen initial
condition the only finite fixed point achieved at large $|t|$  is
$g_c^{(3)}$ for which all directions are attractive with eigenvalues
$x=\{2,1,1/6\}$ at the fixed point. For all the rest fixed points the
probability to appear exactly at the attractive domains of 2 or 3
dimensions in the whole 3 or 4 dimensional volume is neglegible.
Exceptions are related to choosing the initial conditions precisely
at these attractive domains. This occurs with $g_c^{(4)}$ when one takes
$g_2=0$ and with $g_c^{(1)}$ when additionally $g_3=0$ and one chooses
the initial points lying in the attractive domains (see Sections 6.3 and
6.6). We checked this circumstance by repeating our evolution  exluding 
from the initial points (\ref{inidom}) values $g_2=0$ or $g_2=g_3=0$.
Then in the first case one finds convergence to fixed points $g_c^{(3)}$
and $g^{(4)}$ with the number of events $105006:8400$,  respectively,
out of the total ones  176400. In the second case only  point
$g_c^{(3)}$ survives in the limit $|t|>>1$ (the same number of
converging events 105006).

\begin{figure}[h!p]
\begin{center}
\includegraphics[width=8 cm]{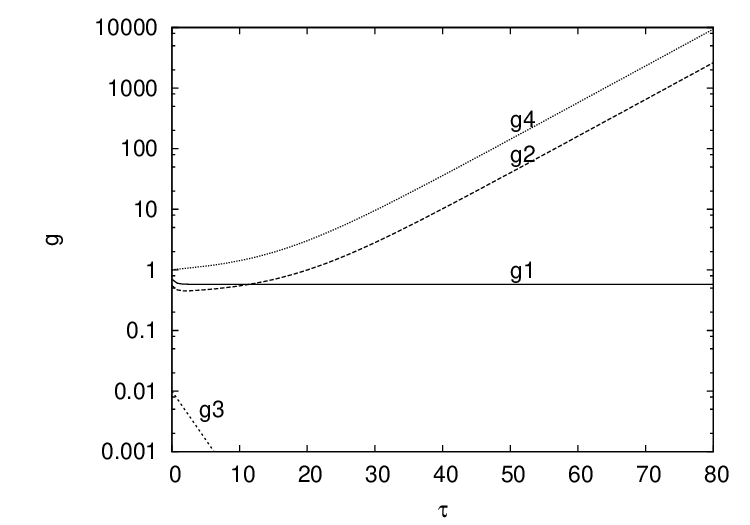}
\includegraphics[width=8 cm]{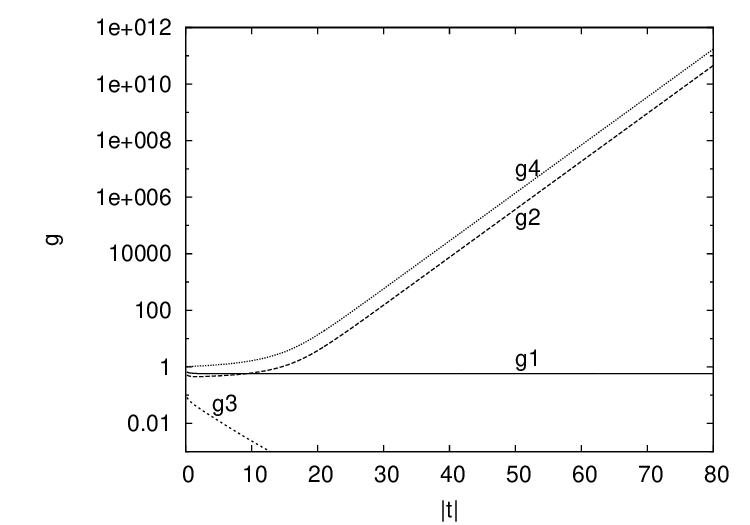}
\caption{Evolution of the coupling constants in the vicinity of the fixed point $g_c^{(2)}$
as $\tau=- t/2$ grows in the linear approximation (left panel) and according to (\ref{evol}) (right panel).
In both cases the constants grow  with $|t|$  (no convergence).}
\label{fig7}
\end{center}
\end{figure}

Note that  one cannot fully guarantee that the stability domain does not
change during evolution. One can discover that such a change may occur 
considering an illustrative example when the initial condition is taken
in the attractive domain for the generally unstable fixed point
$g_c^{(2)}$. Taking some arbitrary initial conditions close to
$g_c^{(2)}$ we first  compared approximate linear evolution with the
full one. We choose
\beq
g=\{0.7,\ 0.55,\ 0.1,\  1\}.
\label{ini2}
\eeq
Recall that
\[g_c^{(2)}=\{0.5774,\ 0.3975,\ 0,\ 0.8896\}.\]
So we expect that the linear evolution should be not so bad.
We also recall that one of the eigenvalues of the stability matrix
at $g_c^{(2)}$ is negative. So we expect that both linear and full
evolution will bring the trajectory away from the fixed point.
The results  shown in Fig.~\ref{fig7} confirm this  behaviour.

\begin{figure}[h!p]
\begin{center}
\includegraphics[width=8 cm]{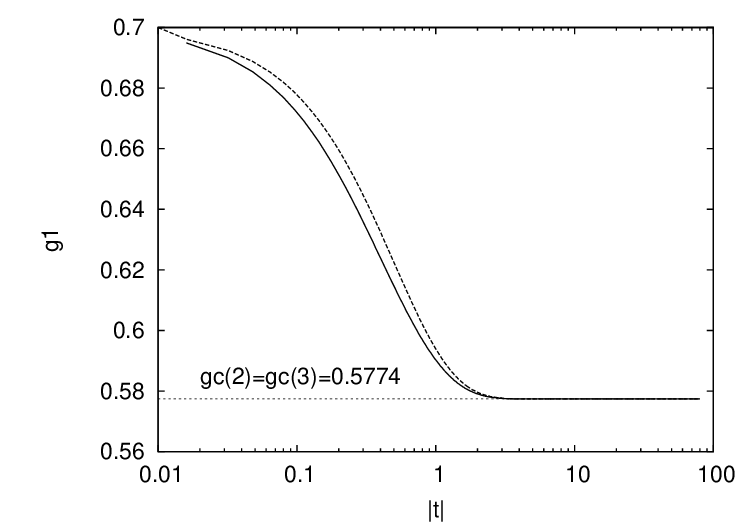}
\includegraphics[width=8 cm]{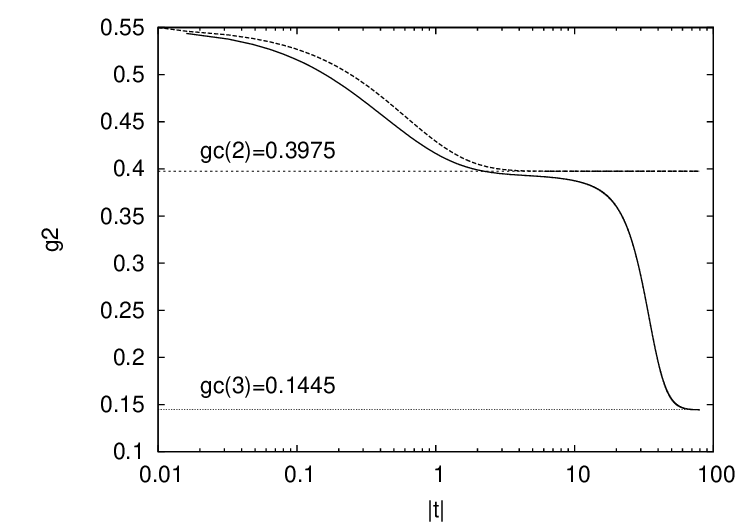}
\includegraphics[width=8 cm]{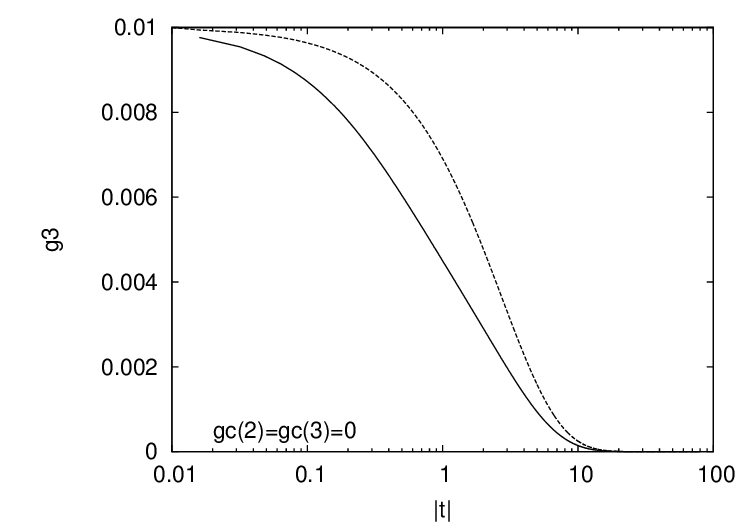}
\includegraphics[width=8 cm]{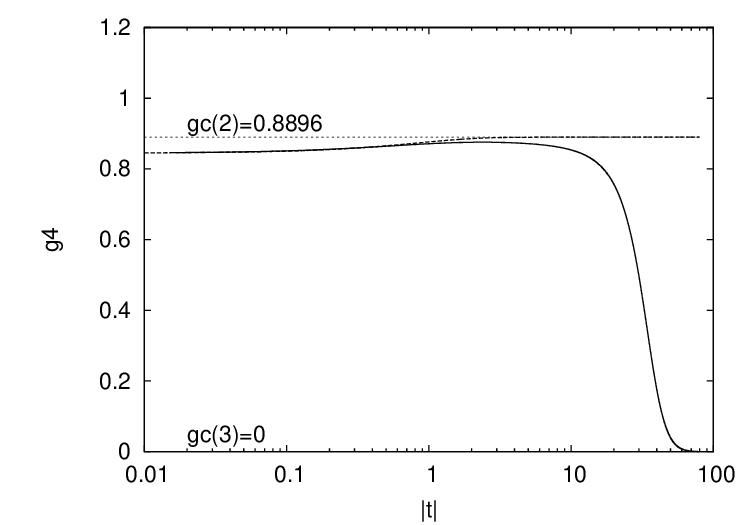}
\caption{Evolution of the coupling constants from the initial data (\ref{ini3})
adjusted to exclude the repulsive direction at $g_c^{(2)}$
via full evolution (\ref{evol}) (solid lines) and in the linear approximation (dashed lines).
Horizontal lines mark values of components of $g_c^{(2)}$ and $g_c^{(3)}$.}
\label{fig8}
\end{center}
\end{figure}

Now we proceed to exclude the repulsive direction at the start adjusting the initial conditions
as described previously. As a result the initial $g_4$ is lowered, so
the adjusted initial condition is
\beq
g_{adj}=\{0.7,\ 0.55,\ 0.1\ ,0.8451\}.
\label{ini3}
\eeq
Both linear and full evolution  with thus adjusted initial condition
is shown in Fig.~\ref{fig8}. We showed the behavior of each constant
$g_1,...,g_4$ separately in $|t|$ variable (in a logarithmical scale)
to see the change occurring at small $|t|$.

With the linear evolution and purely attractive initial domain one
observes an abrupt change from the initial values to the ones tending
to the fixed point $g_c^{(2)}$ at which they stabilize at large rapidities.
However, during full evolution the stability matrix becomes changed,
so that at a certain (rather small) rapidity it restores the repulsive
direction after which $g_2$ and $g_4$ start going away from $g_c^{(2)}$
and begin  going instead to the only stable point $g_c^{(3)}$.
In particular, one sees from Fig.~\ref{fig8} that in the linear
approximation $g_4$ steadily attracts  from 0.8451 to the fixed point
value 0.8896, while in the full evolution at approximately $|t|=1.2$
this initial attraction changes to repulsion, which pushes $g_4$ down
to its value zero at $g_c^{(3)}$. The similar change occurs at $|t|=1.2$
with coupling constant $g_2$. So, obviously, the stability matrix
initially purely attractive acquires  a repulsive direction at this
value of $t$. So in the end full evolution brings the coupling constant
from (\ref{ini3}) to $g_c^{(3)}$. The fixed point $g_c^{(2)}$ to which 
the linear evolution is attracted disappears and is changed
for the stable $g_c^{(3)}$.

\section{Conclusions}

We have studied  the
Regge-Gribov model with two reggeons, pomeron and odderon, in physical transverse dimensions using the RG technique.
The two  intercepts were taken equal to unity from the start following the results found earlier without odderon
~\cite{abarb2}. Presence of the odderon complicates the behaviour of the coupling constants considerably.
Instead of a single attractive fixed point there appear several ones and nearly all fixed points have repulsive directions.
At the fixed points the asymptotic of the propagators is qualitatively the same as without odderon, that is modulated by the
logarithms of energy in certain rational powers. Away from the fixed points the asymptotic strongly depends on the concrete choice
of initial coupling constants. Roughly in third of the cases the latter grow to infinity meaning that our single loop
approximation is inadequate. In the rest cases the coupling constants tend to one of the three fixed points, predominantly
to $g_c^{(3)}$, which is the only one purely attractive. The other two $g_c^{(1)}$ and $g_c^{(4)}$  appear when the initial conditions
are suitably chosen in their respective attractive domains inside the total 4-dimensional volume.
The fixed point $g_c^{(2)}$ found in ~\cite{bartels}, apart from the trivial one, disappears in the course
of evolution in all studied cases because the initial conditions do not exactly fall in its initial attractive domain.

For the five cases of the finite fixed points
we find only two variants of the cross-section high-energy behaviour:
growing as $(\ln s)^{1/6}$ or decreasing as $(\ln s)^{-1}$.

We were not able to rigorously predict when the flows end at some  fixed point
or go to infinity, leaving this task for future studies.
In future we also plan to consider the asymptotical behaviour of the scattering amplitudes. This will require
introducing relevant impact factors and discussion of their possible behaviour with energies and sorts of reggeons.
Our preliminary prediction is that at high energies the amplitudes will be approximately given by the coupling to
the single reggeon propagators, as found without odderon in ~\cite{abarb1}.

\section{Acknowledgements}

The authors are grateful to N.V. Antonov, N.M. Gulitskiy and P.I. Kakin
for very useful discussions.


\newpage
\section{Appendix A. Calculation of the beta-functions}.
\subsection{Derivation}

The renormalized coupling constants are defined by renormalization conditions
\[\Gamma_R^{10,20}\Big|_{r.p}=\frac{\lambda_1}{(2\pi)^{(D+1)/2}},\ \
\Gamma_R^{01,11}\Big|_{r.p}=\frac{\lambda_2}{(2\pi)^{(D+1)/2}},\ \
\Gamma_R^{10,02}\Big|_{r.p}=\frac{\lambda_3}{(2\pi)^{(D+1)/2}}\]
with the renormalization point r.p.=$\{E_1=-E_N.\ E_2=E_3=-E_N/2,\ \  k_i=0\}.$

Generally
\[\Gamma_R^{10,20}=Z_1^{3/2}\Gamma^{10,20}=Z_1^{3/2}\frac{\lambda_{10}}{(2\pi)^{(D+1)/2}}+\Lambda^{10,20}(\lambda.\alpha'),\]
\[\Gamma_R^{01,11}=Z_1^{1/2}Z_2\Gamma^{01,11}=Z_1^{1/2}Z_2\frac{\lambda_{20}}{(2\pi)^{(D+1)/2}}+\Lambda^{01,11}(\lambda,\alpha'),\]
\[\Gamma_R^{10,02}=Z_1^{1/2}Z_2\Gamma^{10,02}=Z_1^{1/2}Z_2\frac{\lambda_{30}}{(2\pi)^{(D+1)/2}}+\Lambda^{10,02}(\lambda,\alpha').\]
Here $\Lambda$ is a sum of all non-trivial (loop) diagrams. In the lowest order one can drop factors $Z$ in front of it and substitute
unrenormalized parameters by the renormalized ones in its arguments.
So at the renormalization point,
 multiplying by $(2\pi)^{(D+1)/2}$
\[\lambda_1=Z_1^{3/2}\lambda_{10}+(2\pi)^{(D+1)/2}\Lambda^{10,20}(\lambda,\alpha')\Big|_{r.p},\]
\[\lambda_2=Z_1^{1/2}Z_2\lambda_{20}+(2\pi)^{(D+1)/2}\Lambda^{01,11}(\lambda,\alpha')\Big|_{r.p},\]
\[\lambda_3=Z_1^{1/2}Z_2\lambda_{30}+(2\pi)^{(D+1)/2}\Lambda^{10,02}(\lambda.,\alpha')\Big|_{r.p}.\]

The $\beta$-function is determined as the derivative in $E_N$ of the dimensionless constants $g_i$, $i=1,2,3$:
\[ g_i=\frac{\lambda_i}{(8\pi\alpha'_1)^{D/4}}E_N^{D/4-1},\ \ \beta_i(g)=E_N\frac{\pd}{\pd E_N}g_i, \ \ i=1,2,3.\]
Differentiation of $g$ gives three terms
\[E_N\frac{\pd}{\pd E_N}g_i=\sum_{j=1}^3t_j,\ \ {\rm for\ each}\ \ i=1,2,3,\]
\[t_1=\frac{\lambda_i}{(8\pi\alpha'_1)^{D/4}}E_N\frac{\pd}{\pd E_N}E_N^{D/4-1}=(D/4-1)g_i,\]
\[t_2=\frac{\lambda_i}{(8\pi)^{D/4}}E_N^{D/4-1}E_N\frac{\pd}{\pd E_N}\frac{1}{{\alpha'_1}^{D/4}},\ \
t_3=\frac{1}{(8\pi\alpha'_1)^{D/4}}E_N^{D/4-1}E_N\frac{\pd}{\pd E_N}\lambda_i.\]
To find $t_2$ we consider
\[E_N\frac{\pd}{\pd E_N}{\alpha'_1}^{-D/4}=
E_N\frac{\pd}{\pd E_N}e^{-(D/4)\ln\alpha'_1}=-(D/4)e^{-(D/4)\ln\alpha'_1}E_N\frac{\pd}{\pd E_N}\ln\alpha'_1=
-(D/4)\tau_1{\alpha'_1}^{-D/4},\]
so
\[t_2=-(D/4)\tau_1 g_i.\]

Term $t_3$ splits into two parts, the first $t_{31}$ from $\lambda_{i0}$ multiplied by the appropriate renormalization constant
and the second $t_{32}$ from $\Lambda$. Both are different for $i=1,2,3$.

Consider $\beta_1$. Then
\[t_{31}=\frac{\lambda_{10}}{(8\pi\alpha'_1)^{D/4}}E_N^{D/4-1}E_N\frac{\pd}{\pd E_N}Z_1^{3/2},\]
\[E_N\frac{\pd}{\pd E_N}Z_1^{3/2}=E_N\frac{\pd}{\pd E_N}e^{(3/2)\ln Z_1}=
e^{(3/2)\ln Z_1}\frac{3}{2}E_N\frac{\pd}{\pd E_N}\ln Z_1=\frac{3}{2}\gamma_1 Z_1^{3/2}\simeq \frac{3}{2}\gamma_1.\]
Here we drop terms of the higher order than considered. So as a result
\[t_{31}=\frac{3}{2}\gamma_1\frac{\lambda_{10}}{(8\pi\alpha'_1)^{D/4}}E_N^{D/4-1}=\frac{3}{2}\gamma_1g_1,\]
where we again dropped higher order terms.
Term $t_{32}$ requires calculations of the non-trivial vertices:
\[t_{32}=\frac{1}{(8\pi\alpha'_1)^{D/4}}E_N^{D/4-1}(2\pi)^{(D+1)/2}E_N\frac{\pd}{\pd E_N}\Lambda^{10,20}(\lambda,\alpha')\Big|_{r.p}.\]

This is repeated for $\beta_2$ and $\beta_3$

For $\beta_2$ we get
\[t_{31}=\Big(\frac{1}{2}\gamma_1+\gamma_2\Big)g_2,\]
\[t_{32}=\frac{1}{(8\pi\alpha'_1)^{D/4}}E_N^{D/4-1}(2\pi)^{(D+1)/2}E_N\frac{\pd}{\pd E_N}\Lambda^{01,11}(\lambda,\alpha')\Big|_{r.p}\]
and for $\beta_3$
\[t_{31}=\Big(\frac{1}{2}\gamma_1+\gamma_2\Big)g_3,\]
\[t_{32}=\frac{1}{(8\pi\alpha'_1)^{D/4}}E_N^{D/4-1}(2\pi)^{(D+1)/2}E_N\frac{\pd}{\pd E_N}\Lambda^{10,02}(\lambda,\alpha')\Big|_{r.p}.\]

To conclude we calculate $\beta_4$ associated with evolution of $g_4\equiv u=\alpha'_2/\alpha'_1$
\[\beta_4=E_N\frac{\pd}{\pd E_N}u.\]
We have seen that
\[ u=u_0\frac{Z_2U_2^{-1}}{Z_1U_1^{-1}}.\]
So
\[\beta_4=u_0E_N\frac{\pd}{\pd E_N}e^{\ln(Z_2U_2^{-1})-\ln(Z_1U_1^{-1})}=
uE_N\frac{\pd}{\pd E_N}(\ln(Z_2U_2^{-1})-\ln(Z_1U_!^{-1})=u(\tau_2-\tau_1).\]

\subsection{The triangular diagram}

\begin{figure}[h!p]
\begin{center}
\includegraphics[width=10 cm]{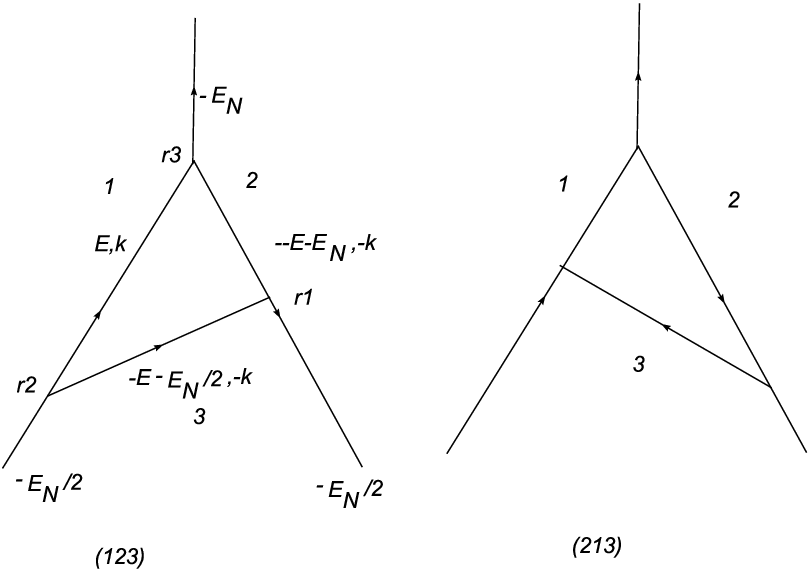}
\caption{The general triangle diagram $\Lambda_{123}$
on the left  and its pair diagram $\Lambda_{213}$ on the right  at renormalzation point.}
\label{figd1}
\end{center}
\end{figure}

We consider a general triangular diagram in which parameters of reggeons 1,2 and 3 may be any of the sets
$\lambda=\{\lambda_1,\lambda_2,\lambda_3\}$ and $\alpha'=\{\alpha'_1,\alpha'_2\}$. This diagram
 is shown in Fig.~\ref{figd1} together with its pair diagram
with the opposite direction of the rung,
the ''inverse'' diagram. We denote the left diagram as $\Lambda_{123}$. After reflection the inverse diagram
acquires the same structure  with the change $1\leftrightarrow 2$ and of
some coupling constants. We denote it as $\Lambda_{213}$.

Following the rules of ~\cite{abarb1} at the renormalization point
\[E_1=-E_N,\ \ E_2=E_3=-E_N/2,\ \  k_1=k_2=k_3=0\]
we have
\[\Lambda_{123}=i^3\frac{\lambda_1\lambda_2\lambda_3}{(2\pi)^{3(D+1)/2}}\]\[\times\int dEd^Dk
(E-\alpha'_1k^2+i0)^{-1}(-E-E_N-\alpha'_2k^2+i0)^{-1}(-E-E_N/2-\alpha'_3k^2+i0)^{-1}.\]
Integration over $E$ gives factor $(-2\pi i)$ and puts $E=\alpha'_1k^2$.
So we get
\[\Lambda_{123}=-\frac{2\pi \lambda_1\lambda_2\lambda_3}{(2\pi)^{3(D+1)/2}}\int d^Dk
(-E_N-(\alpha'_1+\alpha'_2)k^2)^{-1}(-E_N/2-(\alpha'_1+\alpha'_3)k^2)^{-1}.\]

We define
\[\alpha_{12}=\alpha'_1+\alpha'_2,\ \ \alpha_{13}=\alpha'_1+\alpha'_3.\]
Denoting the integral over $k$ as $I$ and using the Feynman parametrization we obtain
\[I=\int_0^1dx\int d^Dk[x(-E_N-\alpha_{12}k^2)+(1-x)(-E_N/2-\alpha_{13}k^2)]^{-2}.
=\int_0^1\frac{dx}{\alpha^2}\int\frac {d^Dk}{(k^2+a^2)^2}\]\[=
\frac{(2\pi)^D}{(4\pi)^{D/2}}\Gamma(2-D/2)(a^2)^{D/2-2},\]
where
\[a^2=\frac{E_N(1+x)/2}{\alpha},\ \ \alpha=x\alpha_{12}+(1-x)\alpha_{13}=\alpha'_1+x\alpha'_2+(1-x)\alpha'_3.\].

So at this point we find the diagram as an integral over $x$
\[-\Lambda_{123}=C\Gamma(2-D/2)\int_0^1\frac{dx(a^2)^{D/2-2}}{\alpha^2}
=-C\Gamma(2-D/2)E_N^{D/2-2}\int_0^1dx[(1+x)/2)]^{D/2-2}\alpha^{-D/2},\]
where the coefficient is
\[C=\lambda_1\lambda_2\lambda_32^{-D/2}(2\pi)^{-D-1/2}.\]
By itself $\Lambda_{123}$ diverges at $D\to 4$ due to presence of $\Gamma(2-D/2)$.
However, in fact we need
\[E_N\frac{\pd}{\pd E_N}\Lambda_{123}=(D/2-2)\Lambda_{123}.\]
Since
\[(D/2-2)\Gamma(2-D/2)=-(2-D/2)\Gamma(2-D/2)=-\Gamma(3-D/2),\]
this derivative actually makes the contribution finite at $\epsilon=0$.
So in the end putting $D=4$ everywhere
\[E_N\frac{\pd}{\pd E_N}\Lambda_{123}=C\int_0^1\frac{dx}{\alpha^2}\ ,\]
where in $C$ one  has to also put $D=4$ and doing the integral over $x$
we get
\beq
E_N\frac{\pd}{\pd E_N}\Lambda_{123}=C\frac{1}{\alpha_{12}\alpha_{13}}=-C\frac{1}{(\alpha'_1+\alpha'_2)(\alpha'_1+\alpha'_3)}.
\label{eqd1}
\eeq
So from the diagram $\Lambda_{123}$ we find the contribution
\[t_{32}=-CE_N^{D/4-1}(2\pi)^{(D+1)/2}\frac{1}{\alpha_{12}\alpha_{13}}\]
Expressing  the product $\lambda_1\lambda_2\lambda_3$ in $C$ by the product $g_1g_2g_3$ of the constants which appear in the diagram
we  obtain
from the diagram $\Lambda_{123}$
\beq
t_{32}=4g_1g_2g_3\frac{{\alpha_1'}^2}{\talpha_{12}\talpha_{13}},
\label{eqd3}
\eeq
where we have redenoted as $\talpha_i$ the slopes in the diagram to distinguish them from $\alpha'_1$
coming from the expression of $\lambda_i$ via $g_i$.
Recall that one has to additionally take into account the inverse diagram $\Lambda_{213}$ in which the reggeons may be
different, so that both $\alpha'_i$, $i=1,2$ and $\tilde{g}_i$ $i=1,2,3$ may be different.

Also the necessary ingredients for the calculation
of the $\beta$-functions are the anomalous dimensions given by
(\ref{gamma1}), (\ref{gamma2}) and (\ref{tau1}), (\ref{tau2}),
where D=4 is substituted.

\subsection{$\beta_4$}
We immediately find
\beq
\beta_4=u(\tau_2-\tau_1)=
\frac{u}{4}g_1^2-\frac{4u^2}{(1+u)^3}g_2^2-\frac{2-u}{4u}g_3^2.
\label{beta4}
\eeq

\begin{figure}[h!p]
\begin{center}
\includegraphics[width=10 cm]{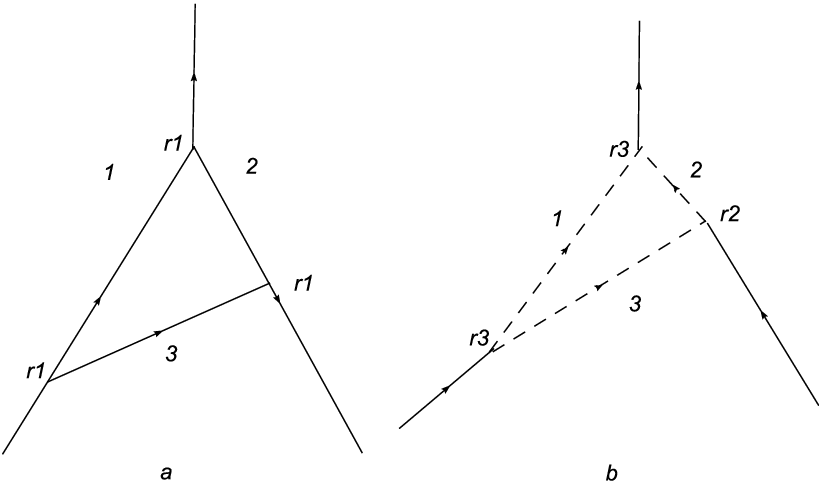}
\caption{Diagrams for $\Lambda^{10,20}$. Pomerons and odderons are shown by solid and dashed lines respectively.
Inverse diagrams are identical.}
\label{figd2}
\end{center}
\end{figure}

\subsection{$\beta_1$}.
We find successively
\[t_1=-\frac{\epsilon}{4}g_1,\ \
t_2=-\tau_1 g_1=g_1\Big(\frac{1}{4}g_1^2-\frac{2-u}{4u^2}g_3^2\Big),\ \
t_{31}=\frac{3}{2}\gamma_1g_1=\frac{3}{2}g_1\Big(-\frac{1}{2}g_1^2+\frac{1}{2u^2}g_3^2\Big).\]
From the diagrams $a$ and $b$ in Fig.~\ref{figd2} and their inverse diagrams we get respectively
\[t_{32}^{a+a'}=2g_1^2,\ \ t_{32}^{b+b'}=-\frac{2}{u^2}g_2g_3^2.\]
Note that  terms with $g_3^2$ should be taken with the extra minus sign.
Summing all terms we find
\beq
\beta_1=-\frac{\epsilon}{4}g_1+\frac{3}{2}g_1^3+g_1g_3^2\frac{1+u}{4u^2}-g_2g_3^2\frac{2}{u^2}.
\label{beta1}
\eeq

\begin{figure}[h!p]
\begin{center}
\includegraphics[width=10 cm]{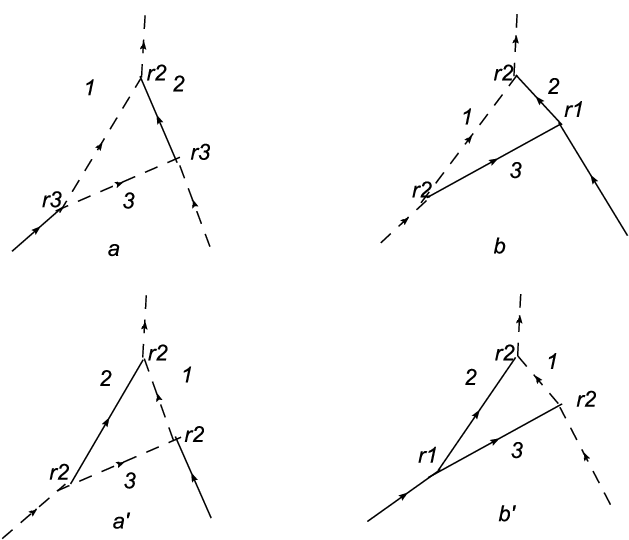}
\caption{Diagrams for $\Lambda^{01,11}$. Pomerons and odderons are shown by solid and dashed lines respectively.
 Inverse diagrams are different and shown below as (a') and (b').}
\label{figd3}
\end{center}
\end{figure}

\begin{figure}[h!p]
\begin{center}
\includegraphics[width=10 cm]{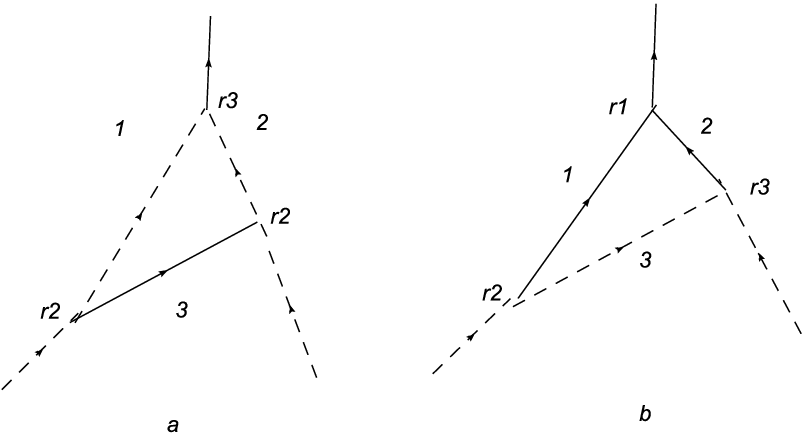}
\caption{Diagrams for $\Lambda^{10,02}$. Pomerons and odderons are shown by solid and dashed lines respectively.
 Inverse diagrams are identical.}
\label{figd4}
\end{center}
\end{figure}

\subsection{$\beta_2$}.
We find successively
\[t_1=-\frac{\epsilon}{4}g_2,\ \
t_2=-\tau_1 g_1=g_2\Big(\frac{1}{4}g_1^2-\frac{2-u}{4u^2}g_3^2\Big),\]
\[t_{31}=\Big(\frac{1}{2}\gamma_1-\gamma_2\Big)g_2=g_2\Big(-\frac{1}{4}g_1^2+\frac{1}{4u^2}g_3^2-\frac{4}{(1+u)^2}g_2^2\Big).\]
From the diagrams $a$ and $b$ in Fig.~\ref{figd3} and their inverse diagrams  we get respectively
\[t_{32}^{a+a'}=-\frac{4}{1+u}\Big(\frac{1}{2u}+\frac{1}{1+u}\Big)g_2g_3^2=-2g_2g_3^2\frac{1+3u}{u(1+u)^2},\]
\[ t_{32}^{b+b'}=-\frac{4}{1+u}\Big(\frac{1}{1+u}+\frac{1}{2}\Big)g_1g_2^2=2g_1g_2^2\frac{3+u}{(1+u)^2}.\]
Terms with $g_1^2g_2$  and $g_2^3$ cancel.
Terms with $g_2g_3^2$ give in the sum
\[
g_2g_3^2\frac{1}{4u^2(1+u)}(-2+u^2-u+1+u-8u)=g_2g_3^2\frac{1}{4u^2(1+u)}(-1+u^2-8u).\]
Adding all the rest terms we find
\beq
\beta_2=-\frac{\epsilon}{4}g_2+g_1g_2^2\frac{6+2u}{(1+u)^2}-g_2g_3^2\frac{1+8u-u^2}{4u^2(1+u)}.
\label{beta2}
\eeq

\subsection{$\beta_3$}.
We find successively
\[t_1=-\frac{\epsilon}{4}g_3,\ \
t_2=-\tau_1 g_1=g_3\Big(\frac{1}{4}g_1^2-\frac{2-u}{4u^2}g_3^2\Big),\]
\[t_{31}=\Big(\frac{1}{2}\gamma_1-\gamma_2\Big)g_3=g_3\Big(-\frac{1}{4}g_1^2+\frac{1}{4u^2}g_3^2-\frac{4}{(1+u)^2}g_2^2\Big).\]
From the diagrams $a$ and $b$ in Fig.~\ref{figd4} and their inverse diagrams  we get respectively
\[t_{32}^{a+a'}=g_2^2g_3\frac{4}{u(1+u)},\ \ t_{32}^{b+b'}=g_1g_2g_3\frac{4}{1+u}.\]

Summing all terms we find
\beq
\beta_3=-\frac{\epsilon}{4}g_3+g_1g_2g_3\frac{4}{1+u}+g_2^3g_3\frac{4}{u(1+u)^2}+g_3^3\frac{u-1}{4u^2}.
\label{beta3}
\eeq

\subsection{Matrix $a_{ik}=\{2\pd\beta_i/\pd g_k\}$}

\[a_{11}=-\frac{1}{2}\epsilon+9g_1^2+ {g}_3^2\frac{1+u}{2u^2},\ \
a_{12}=- {g}_3^2\frac{4}{u^2},\]
\[a_{13}=- 2g_2g_3\frac{4}{u^2}+ 2g_1g_3\frac{1+u}{2u^2},\ \
a_{21}=g_2^2\frac{12+4u}{(1+u)^2},\]
\[a_{22}=-\frac{1}{2}\epsilon+2g_1g_2\frac{12+4u}{(1+u)^2}- g_3^2\frac{1+8u-u^2}{2u^2(1+u)},\ \
a_{23}=- 2g_2g_3\frac{1+8u-u^2}{2u^2(1+u)},\]
\[a_{31}=g_2g_3\frac{8}{1+u},\ \
a_{32}=g_1g_3\frac{8}{1+u}+2g_2g_3\frac{8}{u(1+u)^2},\]
\[a_{33}=-\frac{1}{2}\epsilon+g_1g_2\frac{8}{1+u}+g_2^2\frac{8}{u(1+u)^2}+ 3g_3^2\frac{u-1}{2u^2},\]
\[a_{41}=2g_1\frac{u}{2},\ \
a_{42}=-2g_2\frac{8u^2}{(1+u)^3},\ \
a_{43}= 2g_3\frac{u-2}{2u},\]
\[a_{14}=
+\frac{8}{u^3}g_2g_3^2-\frac{u+2}{2u^3}g_1g_3^2,\]
\[a_{24}=
-\frac{20+4u}{(1+u)^3}g_1g_2^2-\frac{u^3-16u^2-11u-2}{2u^3(1+u)^2}g_2g_3^2,\]
\[a_{34}=
-\frac{8}{(1+u)^2}g_1g_2g_3-\frac{8(1+3u)}{u^2(1+u)^3}g_2^2g_3+ \frac{2-u}{2u^3}g_3^3,\]
\[a_{44}=
\frac{1}{2}g_1^2-\frac{u(16-8u)}{(1+u)^4}g_2^2+\frac{1}{u^2}g_3^2.\]

\section{Appendix B. Real fixed points}
\subsection{$g_3=0$}
Here $r=0$ is assumed.
In this case $\beta_3=0$ and we have three equations
(divided by $(\epsilon/2)^{3/2}$)
\beq
2\beta_1=-g_1+3g_1^3=0,
\label{g31}
\eeq
\beq
2\beta_2=-g_2+g_1g_2^2\frac{12+4u}{(1+u)^2}=0,
\label{g32}
\eeq
\beq
2\beta_4=g_1^2\frac{u}{2}-g_2^2\frac{8u^2}{(1+u)^3}=0.
\label{g33}
\eeq

From (\ref{g31}) we get
\beq
g_1^2=\frac{1}{3},\ \ g_1=\frac{1}{\sqrt{3}}.
\label{g34}
\eeq
Then Eq. (\ref{g32}) gives
\[g_2=0\ \ {\rm or}\ \  -1+g_1g_2\frac{12+4u}{(1+u)^2}=0.\]
If $g_2=0$ then from (\ref{g33}) we get $u=0$.
So our first fixed point is
\[ g_c^{(1)}=\{\frac{1}{\sqrt{3}},\ 0,\ 0,\ 0\} \]
(the pure pomeron model).

If $g_2\neq 0$ then we find
\beq
g_2=\frac{\sqrt{3}(1+u)^2}{12+4u}.
\label{g35}
\eeq
Equation $\beta_4=0$ in its turn admits two solutions.

If $u\neq 0$ then
putting (\ref{g34}) and (\ref{g35}) in (\ref{g33}) we get an equation for $u$
\[ \frac{1}{6}-\frac{24u(1+u)}{(12+4u)^2}=0,\]
which is a quadratic equation
$
8u^2+3u-9=0
\label{g36}
$
with a positive solution
\beq
u=\frac{1}{16}(-3+\sqrt{297})= \frac{3}{16}(\sqrt{33}-1)=0.8896.
\label{g37}
\eeq
With this value of $u$ we find
\beq
g_2=0.3975.
\label{g38}
\eeq
So the second fixed point is
\[ g_c^{(2)}=\{\frac{1}{\sqrt{3}},\ 0.3975,\ 0,\ 0.8896\}. \]

The third fixed point corresponds to $u=0$. Then one finds
\[g_2=\frac{\sqrt{3}}{12},\]
so that
\[ g_c^{(3)}=\{\frac{1}{\sqrt{3}},\ \frac{\sqrt{3}}{12},\ 0,\ 0\}. \]

\subsection{$g_1=g_2=0$}
In this case we have (divided by $(\epsilon/2)^{3/2}$)
\beq
\beta_1=\beta_2=0,\ \ 2\beta_3=-g_3 + g_3^3\frac{u-1}{2u^2},\ \ 2\beta_4=g_3^2\frac{u-2}{2u}.
\label{eq001}
\eeq
So the fixed point  occurs at
\beq
u=2,\ \ g_3= 2\sqrt{2}\sqrt\frac{\epsilon}{2}.
\label{fp00}
\eeq

\end{document}